\documentclass[12pt]{article}
\usepackage{hyperref,url,fullpage,setspace}
\usepackage{amssymb,amsmath,amsthm,amsfonts,mathrsfs,tikz}
\usepackage{graphicx,mathtools,color}
\usepackage{notes2bib}
\usepackage[labelfont=bf]{caption}
\def\di{\mathrm{d}}
\newcommand{\Mpl}{M_\textsf{Pl}}
\usepackage{cite}

\def\C{\mathbb{C}}
\def\R{\mathbb{R}}

\def\zturn{z_\textsf{turn}}
\def\Re{\textsf{Re}}
\def\Im{\textsf{Im}}
\def\HoRe{H_0^\Re}
\def\phioRe{\phi_0^\Re}
\def\phioIm{\phi_0^\Im}
\def\epsoRe{\varepsilon_{1,0}^\Re}

\newcommand\blfootnote[1]{
  \begingroup
  \renewcommand\thefootnote{}\footnote{#1}
  \addtocounter{footnote}{-1}
  \endgroup
}

\usetikzlibrary{shapes.misc}
\tikzset{cross/.style={cross out, draw=black, minimum size=2*(#1-\pgflinewidth), inner sep=0pt, outer sep=0pt},cross/.default={2.5pt}}

\usepackage[intoc]{nomencl}
\makenomenclature

\begin{document}
%%%%%%%%%%%%%%%%%%%%%%%%%%%%%%%%%%%%%%%%%%%%%
\begin{titlepage}
\hfill \\
\vspace*{15mm}
\begin{center}
{\Large \bf Slow-roll approximation in quantum cosmology}

\vspace*{15mm}

{\large Oliver Janssen\blfootnote{e-mail: \href{mailto:oliverjanssen@nyu.edu}{\tt{ojanssen@ictp.it}}}}

\vspace*{8mm}

\small Center for Cosmology and Particle Physics, New York University, New York, NY 10003, USA \\
\small Institute for Fundamental Physics of the Universe and \\
\small International Centre for Theoretical Physics, Trieste 34151, Italy \\

\vspace*{0.7cm}

\end{center}
\begin{abstract}

\noindent In minimally coupled scalar field theories with a potential of the slow-roll type, we give a detailed description of the complex $O(4)$-symmetric solutions to Einstein's equations on the four-ball which contribute to the no-boundary amplitude $\Psi_\textsf{NB}(b,\chi)$ for a closed universe to contain a round three-sphere spatial slice of size $b$ covered homogeneously with the scalar field at value $\chi$. Our derivation demonstrates a result anticipated by Hartle, Hawking and Hertog in \href{https://journals.aps.org/prd/abstract/10.1103/PhysRevD.77.123537}{\textit{Phys. Rev. D} \textbf{77} (2008) 123537}, sharpens Vilenkin's result in \href{https://journals.aps.org/prd/abstract/10.1103/PhysRevD.37.888}{\textit{Phys. Rev. D} \textbf{37} (1988) 888} in the classical regime of the minisuperspace and makes use of a complexified slow-roll approximation. Our technique applies to both the Hartle-Hawking and Vilenkin wave functions, which both predict a family of inflationary universes but weight each member exponentially differently in the semiclassical approximation.

\end{abstract}

\vspace{1.5cm}
\small \today

\end{titlepage}
%%%%%%%%%%%%%%%%%%%%%%%%%%%%%%%%%%%%%%%%%%%%%

%%%%%%%%%%%%%%%%%%%%%%%%%%%%%%%%%%%%%%%%%%%%%
\tableofcontents
%%%%%%%%%%%%%%%%%%%%%%%%%%%%%%%%%%%%%%%%%%%%%

%%%%%%%%%%%%%%%%%%%%%%%%%%%%%%%%%%%%%%%%%%%%%
\nomenclature[01]{$q^\alpha$}{general minisuperspace coordinate describing a homogeneous four-geometry: function of either physical time $t$ (for a classical cosmology) or a complex coordinate $z$ (for an instanton)}
\nomenclature[02]{$\mathfrak{q}^\alpha$}{general minisuperspace coordinate when used as the argument of a wave function (i.e. describing a three-geometry for which we seek the quantum amplitude)}
\nomenclature[03]{$\varphi$}{dimensionful homogeneous scalar field on a four-geometry, function of either $t$ or $z$}
\nomenclature[04]{$a,\phi$}{dimensionless scale factor and homogeneous scalar field when they describe a four-geometry: functions of either $t$ or $z$ depending on the context (relations between dimensionful/dimensionless variables are stated in \S\ref{classicalcosmology})}
\nomenclature[05]{$b,\chi$}{dimensionless scale factor and scalar field when used as the arguments of a wave function; they represent the size of a round three-sphere and the homogeneous value of a scalar that covers it}
\nomenclature[06]{$w^\Re, w^\Im$}{respectively the real and imaginary parts of a complex variable $w$}
\nomenclature[07]{$\phi_0$}{generally complex value of the scalar field at the center of the four-ball on which the no-boundary instantons that we consider are defined}
\nomenclature[08]{$H_0, \varepsilon_{1,0}, \cdots$}{the functions $H(\phi) \equiv \sqrt{V(\phi)/3}, \varepsilon_1(\phi), \cdots$ are to be evaluated at $\phi_0$, i.e, $H(\phi_0), \varepsilon_1(\phi_0), \cdots$}
\nomenclature[09]{$z_1$}{value of the complex coordinate $z$ where the arguments of the wave function are attained, $(a(z_1),\phi(z_1)) = (b,\chi)$}
\nomenclature[10]{$\zturn$}{$z_1^\Im$}

\printnomenclature[2cm]
%%%%%%%%%%%%%%%%%%%%%%%%%%%%%%%%%%%%%%%%%%%%%

%%%%%%%%%%%%%%%%%%%%%%%%%%%%%%%%%%%%%%%%%%%%%
\section{Introduction and conclusion} \label{intro}
%%%%%%%%%%%%%%%%%%%%%%%%%%%%%%%%%%%%%%%%%%%%%
The detailed phenomenology of all but the simplest models of inflation depends significantly on a choice of initial conditions. These models include most multifield theories\footnote{Several works (e.g. \cite{Frazer:2013zoa,Easther:2013rva}) claim the opposite, but their conclusion hinges on the restrictive assumption of sum-separability of the inflaton potential or even simpler: quadratic inflation.} that have been argued to arise from string theory\footnote{For a review see \cite{baumann_mcallister_2015}. The foundations of many string cosmology models are currently being questioned by the swampland program, however. For a review of the state-of-the-art see \cite{Palti:2019pca}.}, for example many-axion theories \cite{Arvanitaki:2009fg} (e.g. \cite{Bachlechner:2018gew}) and D-brane inflation \cite{Dvali:1998pa} (e.g. \cite{McAllister:2012am,Hertog:2015zwh}, but see \cite{Conlon:2019uuy}). But even in the simple scenario of a single scalar field minimally coupled to gravity and subject to a potential with a single inflationary region one can ask basic questions such as what mechanism caused inflation to start in the first place (cf. \cite{East:2015ggf,Kleban:2016sqm,Clough:2016ymm,Aurrekoetxea:2019fhr}) and what determines its duration. Quantum cosmology is an attempt to answer these questions by providing a theory of initial conditions. This happens by modeling the entire universe as a quantum mechanical system described by a wave functional $\Psi$. When, in the semiclassical limit, $\Psi$ takes on a WKB form with a rapidly varying phase compared to its magnitude, it predicts classical evolution and (conditional \cite{HAWKING1986185,Penrose1986-PENQCI}) probabilities for a collection of classical universes may be inferred from (ratios of) the flux of the associated conserved current through surfaces of codimension one in superspace \cite{PhysRev.160.1113,Misner1972,Vilenkin:1988yd,Halliwell:1990uy,Hartle:2018yul} (see also \S\ref{QCsec}), thus effectively providing a measure on initial conditions.

Specifically quantum cosmology deals with closed universes, where, for four spacetime dimensions, $\Psi$ depends on the induced metric and configuration of matter fields on a compact spacelike three-surface $\Sigma$: $\Psi = \Psi[h_{ij},\chi]$. Since the classical theory has at least four constraints -- the consequences of coordinate invariance -- to quantize it one proceeds in canonical quantum gravity by Dirac's procedure \cite{dirac2001lectures} in which physical quantum states are annihilated by operator versions of the constraints \cite{PhysRev.160.1113,Wheeler1968}:
\begin{equation} \label{diraceqns}
	\hat{\mathcal{H}}^\mu \left( \hat{q}, -i \hbar \, \delta/\delta q \right) \Psi = 0 \,,
\end{equation}
where $\mathcal{H}^\mu = 0$ classically and $q$ denotes all the superspace degrees of freedom $\{ h_{ij}(\boldsymbol{\Omega}), \chi(\boldsymbol{\Omega}) \}$ ($\boldsymbol{\Omega}$ is a coordinate that runs over $\Sigma$). One way to solve the functional PDE \eqref{diraceqns} is by a path integral construction of the form
\begin{equation} \label{NBPI}
	\Psi_\textsf{NB}[h_{ij},\chi] = \sum_\mathcal{M} \overset{(g,\phi)|_{\partial \mathcal{M}} = (h,\chi)}{\hspace{-0.75cm} \int} \hspace{-1.1cm} \mathcal{D}g_{\mu \nu} \mathcal{D}\phi \,\, e^{iS/\hbar}
\end{equation}
first put forward by Hartle and Hawking \cite{Hawking1981,PhysRevD.28.2960} (for a discussion of the appropriate action, measure and integration domain so that \eqref{NBPI} indeed solves \eqref{diraceqns} see \cite{PhysRevD.43.1170}, for further constraints see \cite{HarHal1990}), known as the no-boundary (NB) proposal. Vilenkin has made a similar proposal \cite{Vilenkin1982}, at least in one version of it \cite{Vilenkin1984,Vilenkin:2018dch}: the difference will be discussed in \S\ref{NBsec}. Since the proposals are very similar in spirit (though they differ in important details which lead to radically different predictions), we will on occasion call wave functionals of the general form \eqref{NBPI} ``a'' NB wave functional or amplitude (see also \cite{Halliwell:2018ejl}) instead of ``the'' NB wave functional/amplitude. In \eqref{NBPI} the integrated fields live on compact four-manifolds $\mathcal{M}$ with a single boundary $\partial \mathcal{M}$ on which they take the arguments of the wave functional (the class of manifolds in the sum is left undetermined). Though several important elements are left unspecified, this definition has the appealing feature that apart from the requirement that the arguments of the wave functional should be induced on some three-slice, no other boundary conditions (such as what happens ``at the beginning'', because there is none) have to be imposed.

In this paper we will consider the NB amplitudes for a closed universe with three-sphere ($S^3$) spatial topology that contain a round $S^3$ slice of surface area proportional to $b^3$, covered homogeneously with a single minimally coupled scalar field $\varphi$ that takes the value $\chi$ in Planck units:
\begin{equation} \label{PsiNBbchi}
	\Psi_\textsf{NB}[h_{ij}(\boldsymbol{\Omega}) \propto b^2 \Omega_{ij}, \varphi(\boldsymbol{\Omega}) \equiv \Mpl \chi] \equiv \Psi_\textsf{NB}(b,\chi) \,,
\end{equation}
where $\Omega_{ij}$ are the components of the round metric on the unit $S^3$. In this special case the functional PDEs \eqref{diraceqns} reduce to a single ``ordinary'' PDE known as the Wheeler-DeWitt (WDW) equation, $\hat{\mathcal{H}} \Psi = 0$ (see \S\ref{QCsec}). We will restrict ourselves to the leading order approximation of such NB amplitudes in the semiclassical limit (so we will only compute solutions to the classical equations of motion), and assume that the dominant classical configuration that contributes to the path integral lives on a four-ball ($B^4$), is regular and additionally has $O(4)$ symmetry. Further we will be interested in the scenario where the potential energy density $\tilde{V}(\varphi)$ of the scalar is of the slow-roll type, i.e. there is a region where the slow-roll parameters are small over an extended range. Our result, stated in \S\ref{NBsec} in Eqns. \eqref{VilenkinWF}-\eqref{HHWF} and which has been anticipated in Ref. \cite{Hartle:2008ng}, extends the well-known result in the case of a constant positive scalar potential \cite{PhysRevD.28.2960,Halliwell1988} to arbitrary slow-roll models (and generalizes the result in \cite{PhysRevD.46.1546} which studies the particular slow-roll model $\tilde{V}(\varphi) = m^2 \varphi^2$). In the constant-potential case the classical solution, which we will also call a NB ``instanton'' or ``saddle'', can be viewed as half of a four-sphere ($S^4$) (which is responsible for the magnitude of the wave function) glued onto half of de Sitter (dS) space (responsible for the phase). In the general slow-roll case the NB solution is inherently complex, but may be viewed as an approximately Euclidean $S^4$ glued onto approximately Lorentzian dS space (both still essentially responsible for the magnitude and phase of the wave function, respectively). Depending on the choice of contributing saddle, NB wave functions predict a one-parameter family of classical inflating universes which are weighted as
\begin{equation} \label{ourresult}
	|\Psi_\textsf{NB}|^2 \propto \exp \left( \pm \text{constant} \times \frac{\Mpl^4}{\hbar \, \tilde{V}(\varphi_0)} \right)
\end{equation}
where $\varphi_0$ is the starting point of inflation (that is, when the universe had size $a H \approx 1$ and is close to the attractor). As is well-known the sign difference in \eqref{ourresult} distinguishes the Hartle-Hawking from the Vilenkin wave functions (again see \S\ref{NBsec}).

The rest of this paper is organized as follows: in \S\ref{MSPsec} we review homogeneous minisuperspace models where the focus is on very particular slices of the wave functional including the one in Eq. \eqref{PsiNBbchi}. This includes a discussion of the classical cosmology of homogeneous and isotropic spacetimes, the WDW equation and the instanton solution method. In \S\ref{NBsec} we discuss NB instantons, which give rise to particular solutions of the WDW equation, focussing on $O(4)$-symmetric ones on $B^4$. We review the calculation in the case of a constant potential in \S\ref{dSMSP} and turn to slow-roll models -- the main topic of this work -- in \S\ref{SRsec}. In our description of the NB instantons we will use an approximation which extends the usual slow-roll approximation in classical cosmology to quantum cosmology, where the instantons are complex functions. Our new finding is that the equations of motion for all components of the complex fields split in two parts along the approximately Lorentzian part of the solution. The real parts of the fields obey the usual classical slow-roll equations, and so may be solved for separately, while the imaginary parts satisfy their own equations which depend on the real parts. We are able to solve these last equations for the imaginary parts explicitly in terms of the real parts to leading order in the slow-roll parameters, obtaining a detailed approximation of how the imaginary parts decay to zero along the approximately Lorentzian dS phase. This information is vital to a correct estimate of the NB amplitudes \eqref{VilenkinWF}-\eqref{HHWF}, in particular to arrive at Eq. \eqref{ourresult}. In \S\ref{discussionsec} we finish with a discussion containing comments on the regime of validity of our result, the measure on initial conditions for inflation provided by a NB wave function, the existing literature on this topic, the overshoot problem and the extension of our results to multifield models.

%%%%%%%%%%%%%%%%%%%%%%%%%%%%%%%%%%%%%%%%%%%%%
\section{Homogeneous (scalar) minisuperspace models} \label{MSPsec}
%%%%%%%%%%%%%%%%%%%%%%%%%%%%%%%%%%%%%%%%%%%%%
In this section we review some aspects of homogeneous minisuperspace models, focussing on homogeneous scalar minisuperspace models where the degrees of freedom are a scale factor and a homogeneous scalar field. There are many references dealing with this topic, e.g. \cite{Halliwell:1990uy,Halliwell1990,Hartle:2008ng} -- the main purpose here is to set the notation. We will distinguish between two kinds of formulas: those valid for general homogeneous minisuperspace models (where the degrees of freedom are labeled by the letter $q$, or no explicit reference is made to the fields as in Eq. \eqref{WDWeq}) and those valid specifically for homogeneous scalar minisuperspace models (where the degrees of freedom $a,b,\varphi,\phi,\chi$ appear explicitly). Following \cite{PhysRevD.46.1546} we will denote the real and imaginary parts of a complex variable by superscripts $\Re$ and $\Im$ respectively.

%%%%%%%%%%%%%%%%%%%%%%%%%%%%%%%%%%%%%%%%%%%%%
\subsection{Classical cosmology} \label{classicalcosmology}
%%%%%%%%%%%%%%%%%%%%%%%%%%%%%%%%%%%%%%%%%%%%%
The action of a homogeneous scalar 
\begin{equation} \label{rescaledfield}
	\varphi(t) \equiv \Mpl \, \phi(t)
\end{equation}
subject to a potential
\begin{equation}
	\tilde{V}(\varphi) \equiv \mathcal{V}_{S^3} \Mpl^4 ~ V \left( \frac{\varphi}{\Mpl} \right) \notag
\end{equation}
and minimally coupled to the closed FLRW metric
\begin{equation} \label{rescaledmetric}
	g_{\mu \nu} \di x^\mu \di x^\nu \equiv \frac{1}{\mathcal{V}_{S^3} \Mpl^2} \left( -\di t^2 + a(t)^2 \di \boldsymbol{\Omega}_3^2 \right) \,,
\end{equation}
where $\di \boldsymbol{\Omega}_3^2$ is the round metric on the unit $S^3$, $\mathcal{V}_{S^3} = 2 \pi^2$ is its volume and $\Mpl^2 \equiv 1/8 \pi G$, reads\footnote{$\phi, V, t$ and $a$ are all dimensionless, while $[\varphi] = M, [\tilde{V}] = M^4$, and the $t$ and $a$ of the unscaled metric would have $[t] = [a] = L = M^{-1}$. $c \equiv 1$ in our convention.}
\begin{align}
	S &\equiv \int \di^4 x \sqrt{-g} \left( \frac{\Mpl^2}{2} R - \frac{1}{2} (\partial \varphi)^2 - \tilde{V} \right) + \underset{S^3(t_1) \cup S^3(t_2)}{\int} \hspace{-0.6cm} \di \boldsymbol{\Omega}_3 \sqrt{h} \, K \notag \\
	&= \int_{t_1}^{t_2} \di t ~ 3 a \left( 1 - \dot{a}^2 \right) + a^3 \left( \frac{\dot{\phi}^2}{2} - V \right) \notag \\
	&\equiv \int_{t_1}^{t_2} \di t ~ \frac{1}{2} f_{\alpha \beta} \dot{q}^\alpha \dot{q}^\beta - U \equiv \int_{t_1}^{t_2} \di t ~ L(q(t),\dot{q}(t)) \,. \label{generalS}
\end{align}
Here $q \equiv (q^\alpha) \equiv (a,\phi)$ and
\begin{equation}
	(f_{\alpha \beta}) \equiv \begin{pmatrix} 
-6a & 0 \\
0 & a^3
\end{pmatrix} \,, ~~ U \equiv a \left( a^2 V(\phi) - 3 \right) \,. \label{minisupermetric}
\end{equation}
The conjugate momenta are $p_\alpha \equiv \partial L / \partial \dot{q}^\alpha = f_{\alpha \beta} \dot{q}^\beta$, or
\begin{equation} \notag
	p_a = -6a \dot{a} \,, ~ p_\phi = a^3 \dot{\phi} \,.
\end{equation}
The equations of motion (EOM) are
\begin{align}
	\ddot{q}^\alpha + \Gamma^\alpha_{\mu \nu} \dot{q}^\mu \dot{q}^\nu + \nabla^\alpha U &= 0 \,, ~~ \forall \alpha \,, \label{MSPEOM1} \\
	\mathcal{H} \equiv \frac{1}{2} f_{\alpha \beta} \dot{q}^\alpha \dot{q}^\beta + U &= 0 \,, \label{MSPEOM2}
\end{align}
for general homogeneous minisuperspace models\footnote{That is, models described by an action of the form \eqref{generalS}. These can arise more generally e.g. by a metric Ansatz of the type $\di s^2 = - \di t^2 + h_{ij}(q(t)) \di x^i \di x^j$ coupled to homogeneous matter fields.}, where the Christoffel symbols and covariant derivative are with respect to the metric $f$ in Eq. \eqref{minisupermetric}, and
\begin{align}
	\left( \frac{\dot{a}}{a} \right)^2 &= \frac{1}{3} \left( \frac{\dot{\phi}^2}{2} + V \right) - \frac{1}{a^2} \,, \label{FriedmannEQ} \\
	\ddot{\phi} + 3 \frac{\dot{a}}{a} \dot{\phi} + V' &= 0 \,, \label{scalarfieldEQ}
\end{align}
for homogeneous scalar minisuperspace models. To get back to the dimensionful and canonically normalized $\varphi, \tilde{V}$ and an unscaled metric, send
\begin{equation} \notag
	t \rightarrow \sqrt{\mathcal{V}_{S^3}} \Mpl t \,, ~
	a \rightarrow \sqrt{\mathcal{V}_{S^3}} \Mpl a \,, ~
	\phi \rightarrow \frac{\varphi}{\Mpl} \,, ~
	V \rightarrow \frac{\tilde{V}}{\mathcal{V}_{S^3} \Mpl^4} \,.
\end{equation}

%%%%%%%%%%%%%%%%%%%%%%%%%%%%%%%%%%%%%%%%%%%%%
\subsection{Quantum cosmology} \label{QCsec}
%%%%%%%%%%%%%%%%%%%%%%%%%%%%%%%%%%%%%%%%%%%%%
The WDW equation \cite{PhysRev.160.1113,Wheeler1968} in two-dimensional\footnote{More generally the Laplacian should be replaced by the conformal Laplacian \cite{Halliwell:1988wc,Moss:1988wk}.} minisuperspace models is
\begin{equation} \label{WDWeq}
	\hat{\mathcal{H}} \Psi = \left( -\frac{\hbar^2}{2} \nabla^2 + U \right) \Psi = 0 \,.
\end{equation}
As mentioned in \S\ref{intro} we will denote the arguments of the wave function by $(b,\chi)$, which we will abbreviate by $\mathfrak{q}$ for a general homogeneous minisuperspace. The corresponding equation for a WKB state $\Psi \sim P \, e^{i S/\hbar}$ as $\hbar \rightarrow 0$, to leading order in $\hbar$ reads
\begin{equation} \label{HJeq}
	\frac{1}{2} \left( \partial S \right)^2 + U = 0
\end{equation}
generally, or
\begin{equation} \label{HJeqaphi}
	(\partial_b S)^2 - \frac{6}{b^2} (\partial_\chi S)^2 = 12 b^2 \left( b^2 V(\chi) - 3 \right)
\end{equation}
for the scalar minisuperspace. The equation for $P$ is obtained at next-to-leading order in $\hbar$, 
\begin{equation} \notag
	\nabla \cdot \left( P^2 \partial S \right) = 0
\end{equation}
but in this paper we will not discuss this factor in detail.\footnote{Though it may be important to make quantum mechanical sense of certain wave functions, including the NB wave functions which we focus on here (see e.g. \cite{Barvinsky:1990ga,Barvinsky:1992dz}, and the recent \cite{Saad:2019lba,Maldacena:2019cbz,Godet:2020xpk} for exact results in 2D gravity).} When
\begin{equation} \label{classicalitycondition}
	|\partial S^{\Im}|^2 \ll |\partial S^\Re|^2 \,,
\end{equation}
in analogy with non-relativistic quantum mechanics the wave function has been claimed to predict a family of classical universes determined by the integral curves of $S^\Re$ \cite{dirac1930book,Hartle:2008ng,Hartle:2018yul}\footnote{\label{classftnote}In \cite{Hartle:2008ng} the additional conditions $|\partial_\alpha S^{\Im}| \ll |\partial_\alpha S^\Re| \,, \forall \alpha$ are proposed as a proxy for classicality. These conditions are not invariant with respect to minisuperspace coordinate transformations, however, so their meaning is unclear. We believe further investigation into the classical regime in quantum cosmology is warranted (see \cite{PhysRevD.36.3626,PhysRevD.42.585,PhysRevD.42.2566,PhysRevD.42.4056} for earlier discussions on this topic). We thank Thomas Hertog for discussions on this issue.}, that is, the solutions to
\begin{equation} \label{classicaluniverses}
	p_\alpha(q,\dot{q}) = \partial_\alpha S^\Re(q) \,.
\end{equation}
Let us denote a solution by $q_\textsf{cl}(t)$ (to be sure, for the scalar minisuperspace, the classical scalar field four-history and four-metric would be given by Eqns. \eqref{rescaledfield} and \eqref{rescaledmetric} with $(a,\phi) \leftrightarrow (a_\textsf{cl},\phi_\textsf{cl})$). Then $S^\Im(q_\textsf{cl}(t))$ is constant:
\begin{equation} \label{SIconstant}
	\partial_t \, S^\Im(q_\textsf{cl}(t)) = \partial_\alpha S^\Im(q_\textsf{cl}(t)) \, \dot{q}_\textsf{cl}^\alpha(t) = \partial_\alpha S^\Im f^{\alpha \beta} p_{\textsf{cl},\beta} = \partial_\alpha S^\Im f^{\alpha \beta} \partial_\beta S^\Re = 0 \,,
\end{equation}
where $f^{\alpha \beta} \equiv (f^{-1})_{\alpha \beta}$. The final combination in Eq. \eqref{SIconstant} vanishes because of the (imaginary part of) Eq. \eqref{HJeq} for $S(\mathfrak{q})$. We stress that the equations \eqref{classicaluniverses} are \textit{first} order ODEs. If there are $n$ minisuperspace coordinates, the solution space is $(n-1)$-dimensional ($-1$ because of the Hamiltonian constraint). This should be contrasted with the general $(2n-1)$-dimensional solution space to the second order minisuperspace EOM \eqref{MSPEOM1}-\eqref{MSPEOM2}. A WKB wave function satisfying the classicality condition \eqref{classicalitycondition} does not predict just any classical evolution -- it selects a subset \cite{Halliwell:1990uy,Hartle:2018yul}.

Further, one may think of $S^\Im$ as providing a measure on this subset of classical histories via $|\Psi|^2 \approx |P|^2 e^{-2 S^\Im/\hbar}$, since $S^\Im$ is constant along the classical trajectories. More precisely, but not yet fully satisfactory \cite{Halliwell:1992cj}, one works with the conserved current
\begin{equation} \label{JPsi}
	J \equiv - \frac{i \hbar}{2} \left( \Psi^* \nabla \Psi - \Psi \nabla \Psi^* \right) \approx |P|^2 e^{-2 S^\Im/\hbar} \, \nabla S^\Re
\end{equation}
for small $\hbar$, which runs parallel with the classical histories, and considers the flux of this current across codimension-one surfaces in minisuperspace. The relative probability for a classical history to pass through a surface $\Sigma_1$ compared to passing through $\Sigma_2$, and thus to exhibit the properties of those histories passing through $\Sigma_1$ compared to those passing through $\Sigma_2$, is then taken to be the ratio of the fluxes of $J$ through $\Sigma_{1,2}$. These relative probabilities are well-defined because $\nabla \cdot J = 0$. We refer the reader to \cite{Vilenkin:1988yd} and references therein for more details on this general procedure, including a discussion of the caution one must take with negative probabilities arising from the indefinite signature of the metric \eqref{minisupermetric}, and to \cite{Halliwell:2009rw} for a discussion on how the heuristic interpretation sketched above could arise from a rigorous operator formalism. For an example in the specific case of biaxial Bianchi IX minisuperspace we refer the reader to \cite{Janssen:2019sex}.

We now turn to solving the PDE \eqref{HJeq}. The general solution space contains arbitrary functions -- it is infinite-dimensional and generally will not exhibit a classical regime \eqref{classicalitycondition} we are most interested in. One way to systematically select particular solutions, which we will focus on in this paper, is by realizing that Eq. \eqref{HJeq} is satisfied by the action $S(\mathfrak{q})$ of an ``instanton" $q(z)$ -- that is, a (generally complex, say analytic) solution to \eqref{MSPEOM1}-\eqref{MSPEOM2} -- which attains the (real) value $\mathfrak{q}$ at some $z_1(\mathfrak{q})$ and has zero ``energy''. This is the case because we have
\begin{equation} \label{SqNB}
	S(\mathfrak{q}) \equiv S[q(z)] = \int_{\mathcal{C}: 0 \rightarrow z_1(\mathfrak{q})} \hspace{-1cm} \di z \hspace{0.6cm} \frac{1}{2} f_{\alpha \beta}(q) (q')^\alpha (q')^\beta - U(q) = \int_0^1 \di r ~ z_1 \, L(q,\frac{\dot{q}}{z_1}) \,,
\end{equation}
where here $\phantom{x}' \equiv \di / \di z$, $\dot{\phantom{x}} \equiv \di / \di r$, $r \equiv z/z_1(\mathfrak{q})$, and we started the integration at the arbitrary point $z = 0 = r$. Then we compute
\begin{equation} \notag
	\partial_\alpha S = \left[ p_\beta \, \partial_\alpha q^\beta \right]^1_0 + \int_0^1 \di r \, z_1 \partial_\alpha q^\beta \left[ \partial_{q^\beta} L - \partial_z \left( \partial_{(q')^\beta} L \right) \right] - \mathcal{H} \, \partial_\alpha z_1 = p_\alpha \,.
\end{equation}
(We used the EOM and the Hamiltonian constraint, and in this formula $\partial_\alpha \equiv \partial/\partial \mathfrak{q}^\alpha$.) Eq. \eqref{HJeq} follows from this and $\mathcal{H} = 0$.

Notice that for a given instanton $q(z)$ which attains $\mathfrak{q}$ at $z = z_1$, which we will denote by $[q(z), z_1]$, we can immediately construct three others, so that the instanton solution subset to Eq. \eqref{HJeq} is generally four-fold degenerate. Specifically, the couples
\begin{equation} \label{NBinstantons}
	\left[ q(z^*)^*, z_1^* \right] \,, ~~ \left[ q(-z), -z_1 \right] \,, ~~ \left[ q(-z^*)^*, -z_1^* \right]
\end{equation}
also satisfy the EOM, are equally analytic, and attain the real values $\mathfrak{q}$ at the points indicated. If the action of $[q(z), z_1]$ is $S$, then the actions of the associated instantons are $S^*, -S$ and $-S^*$ respectively. From Eq. \eqref{HJeq} it follows that if $S$ is a solution, so too are $S^*, -S$ and $-S^*$. Here we have identified the instantons which are responsible for each of these options.

To complete the instanton solution prescription, it remains to select particular solutions to the EOM \eqref{MSPEOM1}-\eqref{MSPEOM2}, which are only ODEs and so require fewer boundary conditions than the PDE \eqref{HJeq}. This brings us to the NB proposal, which we discuss in detail for homogeneous scalar minisuperspace models in the next section and which is the main focus of this paper. A NB wave function for homogeneous minisuperspace models is given by a path integral of the form
\begin{equation} \label{NBMSP}
	\Psi_\textsf{NB}(\mathfrak{q}) \equiv \sum_\mathcal{M} \overset{q|_{\partial \mathcal{M}} = \mathfrak{q}}{\int} \mathcal{D}q^\alpha \, e^{iS/\hbar} \,,
\end{equation}
where the integrated fields $q$ live on compact four-manifolds $\mathcal{M}$ with a single boundary $\partial \mathcal{M}$ on which they assume the arguments $\mathfrak{q}$ of the wave function. As $\hbar \rightarrow 0$ such wave functions presumably take on a (sum of) WKB form(s) indeed, where the action is determined by an instanton in the way we have anticipated above. 

%%%%%%%%%%%%%%%%%%%%%%%%%%%%%%%%%%%%%%%%%%%%%
\section{No-boundary instantons} \label{NBsec}
%%%%%%%%%%%%%%%%%%%%%%%%%%%%%%%%%%%%%%%%%%%%%
A NB instanton is a (generally complex, regular) solution to the EOM for the metric and matter fields which lives on a compact four-manifold that has a single boundary on which the arguments of the wave function are induced. The simplest such instanton for the slice of the wave function we are interested in here -- the boundary being a round $S^3$ of ``radius'' proportional to $b$ which is homogeneously covered with a scalar field that takes the value $\Mpl \chi$, $\Psi(b,\chi)$ -- lives on $B^4$. We will assume that one or more of such instantons provide the dominant contribution to the wave function $\Psi_\textsf{NB}(b,\chi)$ in the semiclassical limit, so that other four-manifolds appearing in the sum \eqref{NBMSP} are irrelevant \bibnote{Which instanton(s) actually determine the leading semiclassical behavior of $\Psi_\textsf{NB}$ depends on which manifolds are included in the sum \eqref{NBPI} and on the domain of integration over metrics and matter fields on each manifold. In \cite{Janssen:2019sex} we computed the actions of two specific NB instantons (in a model with constant dark energy instead of a dynamical scalar), one on $\C \text{P}^2 \setminus B^4$ and the other on $B^4$, and found the former to be subdominant compared to the latter. This gives one data point of motivation for our assumption. On the other hand it is not necessarily true that instantons on $B^4$ always provide the dominant contribution to any slice of the wave function (assuming indeed they contribute at all) -- counter-examples are the squashed $S^3$ amplitudes considered in \cite{Janssen:2019sex}, where a phase transition between NB saddle points on $B^4$ and $\C \text{P}^2 \setminus B^4$ occurs as the boundary sphere is squashed across a critical value. For a related discussion see \cite{PhysRevD.42.2458}.}.

A $B^4$ can be described by a radial coordinate $r \in [0,1]$ and three angles $\boldsymbol{\Omega}_3$ on concentric $S^3$s, $(X^a) \equiv (r,\boldsymbol{\Omega}_3)$, and the simplest i.e. most symmetric NB instanton can be written in the form
\begin{align}
	G_{a b} \, \di X^a \di X^b &\equiv \frac{1}{\mathcal{V}_{S^3} \Mpl^2} \left( N^2 \di r^2 + a(i N r)^2 \di \boldsymbol{\Omega}_3^2 \right) \,, \label{O4metric} \\
	\phi(i N r) \,, \label{O4field}
\end{align}
where $a, \phi : \C \rightarrow \C$ and $N \in \C$ (as in \S\ref{classicalcosmology} $\phi$ is dimensionless and related to the canonically normalized scalar by Eq. \eqref{rescaledfield}). We will further assume that among the instantons on $B^4$, the $O(4)$-symmetric ones of the type \eqref{O4metric}-\eqref{O4field} provide the dominant semiclassical contribution to $\Psi_\textsf{NB}(b,\chi)$. The round $S^3$ boundary where the wave function lives is located at $r \equiv 1$ and the center of the $B^4$ lies at $r \equiv 0$. We will define $z \equiv i N r$ so that the instanton is
\begin{align} \label{NBmetric}
	G_{a b} \, \di X^a \di X^b &= \frac{1}{\mathcal{V}_{S^3} \Mpl^2} \left( - \di z^2 + a(z)^2 \di \boldsymbol{\Omega}_3^2 \right)\,, \notag \\
	\phi(z) \,.
\end{align}
In this notation the boundary is located at $z_1 = i N \in \C$, so that
\begin{equation} \label{boundarydata}
	a(z_1) = b \,, ~~~ \phi(z_1) = \chi \,,
\end{equation}
while the center of the $B^4$ lies at the origin $z = 0$ of the complex $z$-plane. We will assume $a$ and $\phi$ are analytic in an open region containing the origin. The EOM for $(a,\phi)$ are simply Eqns. \eqref{FriedmannEQ}-\eqref{scalarfieldEQ} of \S\ref{classicalcosmology}, but we stress their different interpretation here despite their identical appearance: here $(a(z),\phi(z))$ are complex-valued functions on a compact space $B^4$ (more precisely, the segment $z \in [0,z_1]$ in the complex plane corresponds to the real segment $r \in [0,1]$ on the $B^4$), while in classical cosmology $(a(t),\phi(t))$ are real-valued and live on the non-compact $\R \times S^3$ (or on the cilinder $[t_1,t_2] \times S^3$). The condition of regularity and the EOM imply that $a$ is an odd function of $z$ in a neighborhood of $z = 0$ with $\pm a(z) \sim iz + \mathcal{O}(z^3)$ as $z \rightarrow 0$, and that $\phi$ is an even function of $z$ in the same region with $\phi(z) \sim \phi_0 + \mathcal{O}(z^2)$ as $z \rightarrow 0$ (more precisely, see Eqns. \eqref{aexpansionbig}-\eqref{phiexpansionbig} later on). The two real degrees of freedom in the value $\phi_0 \in \C$ of the scalar at the center of the ball and the two in the complex value $z_1$ match the four real boundary conditions in Eq. \eqref{boundarydata}. So we expect a discrete solution set to this boundary value problem in general.

\begin{figure}
\centering
\begin{tikzpicture}[scale=2]
	\draw[blue!70!white,very thick,fill=green!30!white] (0,0) circle (35pt);
	\draw[black,fill=black] (0,0) circle (0.5pt);
	\node at (0,1.4) {$h_{ij}(\boldsymbol{\Omega})=b^2 \tilde{\Omega}_{ij} = a(z_1)^2 \tilde{\Omega}_{ij}$};
	\node at (0,-1.4) {$\varphi(\boldsymbol{\Omega})/\Mpl \equiv \chi = \phi(z_1)$};
	\node at (0,0.19) {$a=0,\phi=\phi_0$};
	\node at (-1.6,0.0) {$z = z_1$};
	\node at (0,-0.12) {$z=0$};
\end{tikzpicture}
\caption{The simplest NB instanton that ``fills in'' a round $S^3$ (colored blue) of (real) radius proportional to $b$ on which a scalar field homogeneously takes the (real) value $\Mpl \chi$. We use the convention $\tilde{\Omega}_{ij} \equiv \Omega_{ij}/(\mathcal{V}_{S^3} \Mpl^2)$ where $\Omega_{ij}$ are the components of the round metric on the unit $S^3$. The instanton lives on a $B^4$ (colored green) with center located at $r = 0$ (equivalently $z = 0$) and $S^3$ boundary located at $r = 1$ (equivalently $z = z_1$). Inside the $B^4$ the instanton is described by two generally complex functions $a(z)$ and $\phi(z)$. We assume such instantons provide the leading semiclassical approximation to $\Psi_\textsf{NB} \left[ h_{ij},\varphi \right] = \Psi_\textsf{NB}(b,\chi)$.} \label{fig:M1}
\end{figure}
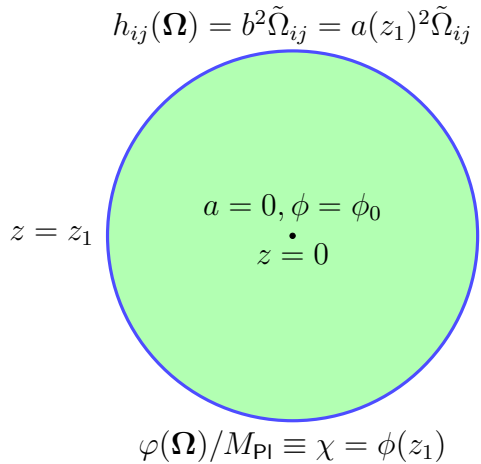

Fig. \ref{fig:M1} depicts an $O(4)$-symmetric NB instanton on $B^4$ and summarizes our conventions. The action of such instantons is as in Eq. \eqref{SqNB}; using the EOM we have
\begin{equation} \label{Sbchicontour}
	S = 2 \int_{\mathcal{C}: 0 \rightarrow z_1} \hspace{-0.7cm} \di z ~~ a(3-a^2 V) \,.
\end{equation}
At this stage we remind the reader of our remark around Eq. \eqref{NBinstantons}, namely that if there is an instanton with action $S$ there are three others related by complex conjugation and parity operations with actions $S^*, -S$ and $-S^*$. This holds in particular for the NB instantons we described above. In this work we will avoid picking a subset of instantons and declaring that these determine the semiclassical wave function. Instead we merely describe the properties of the instantons, leaving the important question of which ones are relevant for the wave function of our universe open. In particular we distinguish the ``no-boundary proposal'' from the ``Hartle-Hawking'' \cite{Hawking1981,PhysRevD.28.2960} wave function and the ``Vilenkin'' \cite{Vilenkin1982,Vilenkin1984,Vilenkin:2018dch} wave function, even though the first two are often identified (but \cite{HarHal1990} makes this distinction too). We take the first term to represent the general idea in Eq. \eqref{NBMSP} -- which is topological, and does not specify the domain of integration over fields $q^\alpha$ -- while the latter two are defined by a specific choice of contributing instantons (cf. \cite{Halliwell:2018ejl} for the Hartle-Hawking wave function), which in turn are determined by the choice of integration domain over the $q^\alpha$.\footnote{We refer the reader to \cite{Feldbrugge:2017kzv,Feldbrugge:2017fcc,Feldbrugge:2017mbc,Feldbrugge:2018gin,DiTucci:2018fdg,DiTucci:2019dji,DiTucci:2019bui,Vilenkin:2018dch,Vilenkin:2018oja,DiazDorronsoro:2017hti,DiazDorronsoro:2018wro,Janssen:2019sex,deAlwis:2018sec} for recent discussions on the NB proposal in minisuperspace models.} We imagine the choice-of-saddles question will be answered by the full theory of quantum gravity in which the wave function of the universe presumably is defined based on a normalization condition that we do not yet understand (which goes beyond the semiclassical reasoning around Eq. \eqref{JPsi}).\footnote{Several consistency conditions on the wave function, e.g. that it must describe a well-defined QFT for small matter fluctuations around the background saddles, have been discussed \cite{HarHal1990}. We are referring here to a normalization condition on the entire wave function, including its behavior on varying backgrounds.}

Before continuing with NB instantons, it might be useful to compare them with Coleman-De Luccia (CDL) \cite{PhysRevD.21.3305} instantons, since both are solutions to the same Eqns. \eqref{FriedmannEQ}-\eqref{scalarfieldEQ} and satisfy seemingly identical initial conditions near $z = 0$. Physically NB instantons are argued to describe the nucleation of the entire universe while CDL instantons describe the decay of an unstable state in a pre-existing universe via the nucleation of a bubble. Mathematically the difference is that CDL instantons are real in the Euclidean direction (i.e. all fields are real, and the metric is of Euclidean signature), while this is not generally the case for NB instantons. In particular $\phi_0 \in \R$ for CDL instantons, while $\phi_0 \in \C$ for NB instantons. Then, in the CDL case, this $\phi_0$ is carefully tuned so that along the Euclidean direction the scale factor $a \rightarrow 0$ in a regular way for a ``second'' time (the ``first'' time being around $z = 0$), while NB instantons instead must attain the values $(b,\chi)$ somewhere in the complex plane (and this will generally be impossible in a purely Euclidean direction). Because of these boundary conditions CDL instantons live on $S^4$ (which has no boundary), while NB instantons live on $B^4$ (which has a boundary).

%%%%%%%%%%%%%%%%%%%%%%%%%%%%%%%%%%%%%%%%%%%%%
\subsection{Constant potential} \label{dSMSP}
\begin{figure}
\centering
\begin{tikzpicture}
      \draw[->,thick] (-4,0) -- (4,0) node[right]{$(H z)^\Re$};
      \draw[->,thick] (0,-4) -- (0,4) node[above]{$(H z)^\Im$};
      \draw (2,-0.1) -- (2,0.1) node[below,yshift=-1.5mm]{$\cosh^{-1}(bH)$};
      \draw (-2,-0.1) -- (-2,0.1) node[below,yshift=-1.5mm]{$-\cosh^{-1}(bH)$};
      \draw (2,1) node[cross]{};
      \draw (2,3) node[cross]{};
      \draw (-2,1) node[cross]{};
      \draw (-2,3) node[cross]{};
      \draw (2,-1) node[cross]{};
      \draw (2,-3) node[cross]{};
      \draw (-2,-1) node[cross]{};
      \draw (-2,-3) node[cross]{};
      \draw (-0.1,-1) -- (0.1,-1) node[right]{$-\pi/2$};
      \draw (-0.1,-3) -- (0.1,-3) node[right]{$-3\pi/2$};
      \draw (-0.1,1) -- (0.1,1) node[right]{$\pi/2$};
      \draw (-0.1,3) -- (0.1,3) node[right]{$3\pi/2$};
      \draw (-5.2,1.03) node[right]{$n=0$};
      \draw (-5.2,3.03) node[right]{$n=1$};
      \draw (-5.2,-0.97) node[right]{$n=-1$};
      \draw (-5.2,-2.97) node[right]{$n=-2$};
      \draw (2,4) node[above]{$\vdots$};
      \draw (-2,4) node[above]{$\vdots$};
      \draw (2,-4) node[above]{$\vdots$};
      \draw (-2,-4) node[above]{$\vdots$};
\end{tikzpicture}
\caption{Information about the $O(4)$-symmetric NB instantons on $B^4$ for the minisuperspace model with constant (positive) vacuum energy density (a.k.a. the ``dS minisuperspace model'' \cite{Halliwell1988}). The crosses denote the the complex coordinates $z_1$ where the real values $(b,\chi)$ -- the arguments of the wave function -- are attained. The functional form of the instantons (distinguished from one another by the endpoint $z_1$) is given in Eqns. \eqref{az}-\eqref{phiz}; the $\pm$ in Eq. \eqref{az} is linked to odd/even values of $|n|$ respectively.} \label{fig:M2}
\end{figure}
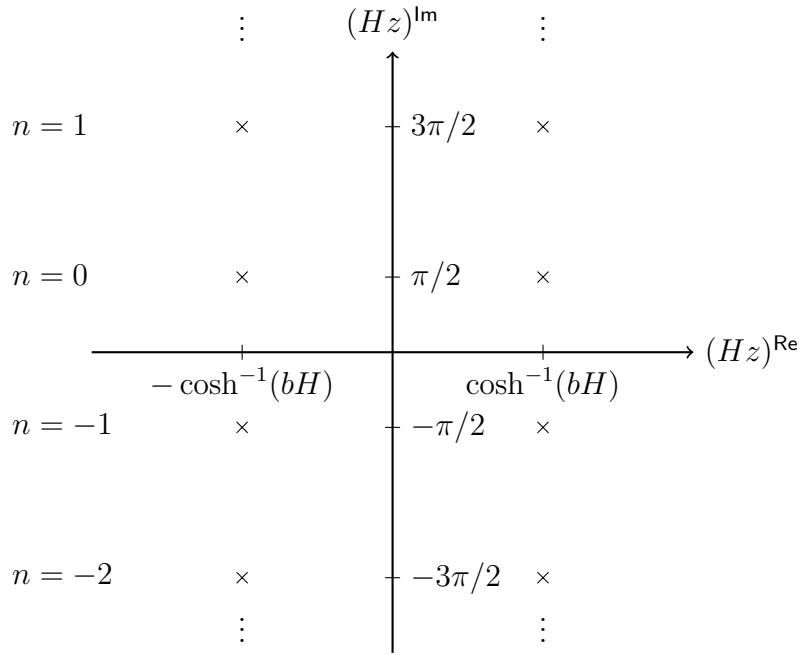

%%%%%%%%%%%%%%%%%%%%%%%%%%%%%%%%%%%%%%%%%%%%%
Before attacking the main problem of this paper, namely the calculation of the NB instantons for scalars subject to a potential with a slow-roll patch, it will be instructive to recall the calculation in the particular case of a constant potential $V(\phi) \equiv 3 H^2$ \cite{PhysRevD.28.2960,Halliwell1988,Hertog:2011ky}. In this case the $O(4)$-symmetric NB instantons on $B^4$ are of the form
\begin{align}
	a(z) &= \pm \frac{1}{H} \sin(i H z) \,, \label{az} \\
	\phi(z) &\equiv \chi \,. \label{phiz}
\end{align}
What distinguishes the instantons is (1) the choice of sign for $a$ and (2) the endpoint $z_1$ where the real values $(b,\chi)$ are attained. In the regime $bH>1$ that we are most interested in, the possible values for $z_1$ are given by
\begin{equation} \label{bH>1}
	H z_1 = \pm \cosh^{-1} (bH) + i \pi \left( n + \frac{1}{2} \right) \,, ~~~ n \in \mathbb{Z} \,.
\end{equation}
For the saddles with the $+$ choice in \eqref{az} $|n|$ must be odd, while for those with the $-$ sign choice $|n|$ must be even. Both signs for the real part in \eqref{bH>1} are allowed for each choice of sign in \eqref{az}; this $\pm$ is not correlated with the $\pm$ in \eqref{az}. This information is displayed in Fig. \ref{fig:M2}.

The actions of these instantons are
\begin{align} \label{iSdSMSP}
	i S &= \left\{ \begin{array}{ll}
       + \displaystyle\frac{2}{H^2} \left( 1 + i \left[ (bH)^2 - 1 \right]^{3/2} \right) & \mbox{for $(H z_1)^\Re < 0$, $|n|$ odd,} \, \vspace{0.2cm} \\
       - \displaystyle\frac{2}{H^2} \left( 1 - i \left[ (bH)^2 - 1 \right]^{3/2} \right) & \mbox{for $(H z_1)^\Re < 0$, $|n|$ even,} \, \vspace{0.2cm} \\
         + \displaystyle\frac{2}{H^2} \left( 1 - i \left[ (bH)^2 - 1 \right]^{3/2} \right) & \mbox{for $(H z_1)^\Re > 0$, $|n|$ odd,} \, \vspace{0.2cm} \\
         - \displaystyle\frac{2}{H^2} \left( 1 + i \left[ (bH)^2 - 1 \right]^{3/2} \right) & \mbox{for $(H z_1)^\Re > 0$, $|n|$ even.} \\
    \end{array} \right.
\end{align}
Notice all the choices of sign for $S^\Re$ and $S^\Im$ appear as we discussed more generally in \S\ref{QCsec}, and that Eq. \eqref{HJeqaphi} is indeed satisfied. These actions were computed via the contour integral \eqref{Sbchicontour}. Since the integrand is an entire function the contour of integration can be chosen to be any curve beginning at the origin and ending at the appropriate $z_1$. For the instantons with $n = 0$ and $n = -1$ there are two famous choices of contour which permit a convenient interpretation of these NB instantons: the first \cite{PhysRevD.28.2960} is a contour which runs vertically from the origin to a ``turning point'' 
\begin{equation} \notag
	H \zturn \equiv \pm \frac{i \pi}{2} \,,
\end{equation}
along which the metric \eqref{NBmetric} is the round metric on an $S^4$ and one hemisphere of this sphere is covered by the segment $[0,\zturn]$, and then runs horizontally to the final value $z_1$, where the metric \eqref{NBmetric} is the metric on dS space in closed slicing, the segment $[\zturn,z_1]$ covering a sequence of spheres from smallest possible (radius $1/H$) to large (radius $ \cosh(bH)/H$). (In particular, $a,\phi$ are real along the entire broken contour.) So these NB instantons can be thought of as half of a (Euclidean) round $S^4$ glued on to part of (Lorentzian) dS space, the equator of the former being glued to the throat of the latter. The vertical or ``Euclidean'' part of the contour determines $(iS)^\Re$ while the horizontal or ``Lorentzian'' part determines $(iS)^\Im$. The second famous choice of contour \cite{Maldacena:2002vr} runs horizontally from the origin to $\pm \cosh^{-1}(bH)$ (or a fraction of this amount), where the metric is \textit{minus} the one on Euclidean anti-dS (with vacuum energy density $-3H^2$), i.e. hyperbolic space. A complex transition is then made to arrive at the value $z_1$. In \S\ref{SRsec} we will consider the adjustments to the first-mentioned representation when the potential is changed from exactly constant to one satisfying the slow-roll conditions. Qualitatively the result is that the NB instanton can be viewed as an approximately Euclidean half-$S^4$ attached to part of an approximately Lorentzian dS space. The dominant contributions to $(iS)^\Re$ and $(iS)^\Im$ come from the approximately Euclidean and Lorentzian pieces respectively. However, the metric is inherently complex in this case (there is no complex diffeomorphism of the $z$-coordinate that renders the metric real along some path in the complex $z$-plane, i.e., it is not a ``real tunneling geometry'' \cite{PhysRevD.42.2458}).

For $n \notin \{-1,0\}$ the instantons can be viewed as multiple complete round $S^4$s plus a hemisphere, glued onto part of dS space. One might have expected the imaginary parts of the actions of these instantons to differ from those with $n \in \{ -1, 0 \}$ by multiples of $4/H^2$, since that is the magnitude of the Euclidean action of a round $S^4$ of radius $1/H$. It turns out though that here a sphere with positive action is always checked by a sphere with negative action\footnote{This feature is peculiar to even dimensions. In e.g. $2 +1$ dimensions the spheres contribute equally and the result is in line with the expectation \cite{PhysRevD.40.4011}.}, so that for each instanton $|S^\Im| = 2/H^2$. The real part of $S$ for the instantons with $n \notin \{-1,0\}$ is more easily seen to remain the same (up to a sign) as those with $n \in \{-1,0\}$.

As we discussed above the Hartle-Hawking and Vilenkin semiclassical wave functions are defined by a choice of saddles in \eqref{iSdSMSP}. The former involves a linear combination of two saddles, one from the first row in \eqref{iSdSMSP} and the other from the third row, while the latter involves just a single instanton from either row two or row four (which one is a convention). In this case both states give the same classical prediction, however: a single dS space of radius $1/H$ with probability one.\footnote{It is sometimes stated (e.g. \cite{Hartle:2008ng}) that the Hartle-Hawking wave function predicts two copies of each classical universe which are time-reversals of each other. ``Time'' here is not the thermodynamic arrow of time, however, but simply a timelike coordinate $t$ which one may redefine at will via e.g. $t \rightarrow -t$. So, assuming they decohere, both contributions to the Hartle-Hawking state describe the same classical universe (the thermodynamic arrow of time runs in the same direction in both).} Since $H$ is constant in this model the factor $e^{\pm 2/H^2}$ that appears in $\Psi_\textsf{NB}$ may be absorbed into the normalization and has no physical consequence. This changes when the scalar is subject to a non-constant potential, as we now describe.

%%%%%%%%%%%%%%%%%%%%%%%%%%%%%%%%%%%%%%%%%%%%%
\subsection{Slow-roll models} \label{SRsec}
%%%%%%%%%%%%%%%%%%%%%%%%%%%%%%%%%%%%%%%%%%%%%
We now turn to the main calculation of interest in this paper, namely the description of the $O(4)$-symmetric NB instantons on $B^4$ in a minisuperspace model where the degrees of freedom are a scale factor and a scalar field subject to a potential with a slow-roll regime.

As reviewed at the beginning of this section, given a $(b,\chi) \in \R^+ \times \R$ which appear in $\Psi_\textsf{NB}(b,\chi)$ we are to search for a complex solution to Eqns. \eqref{FriedmannEQ}-\eqref{scalarfieldEQ} for which \eqref{boundarydata} holds at a point $z_1 \in \C$ and which is regular near the center $z = 0$ of the $B^4$. Even for slow-roll potentials -- as far as the author can tell -- this problem is analytically intractable for \textit{general} values of $(b,\chi)$.\footnote{There are special scalar potentials for which this problem can be solved analytically for all $(b,\chi)$ (see \cite{Garay1990} for a collection, and also \cite{Janssen:2019sex}), but none are of the slow-roll type with the exception of the constant potential.} Instead we will reason the other way round: we will construct particular solutions which attain particular values of $(b,\chi)$ somewhere in the complex plane. By consequence our result for the semiclassical wave function will only be applicable to a limited subset of the minisuperspace (and only for slow-roll potentials). We will return to this point in \S\ref{discussionsec}.
 
 The generalization from constant potential to non-constant potential can be summarized by the generalization of the two quantities $\zturn$ and $\phi_0$ introduced earlier in \S\ref{dSMSP}. In the constant potential case the metric is purely Euclidean (i.e. real, and of Euclidean signature) along the imaginary axis until $\zturn \in i \R$ is reached, after which it is purely Lorentzian (i.e. real, and of Lorentzian signature) along the line $\zturn + t, t \in \R$, and the value $b$ is attained by the scale factor somewhere along this line; $a(\zturn+t_1) = a(z_1) = b$. More generally we will define $\zturn = \Im(z_1)$. Finally in the general case $\R \ni \chi \neq \phi_0 \in \C$ and $\phi(z)$ will be non-constant.

%%%%%%%%%%%%%%%%%%%%%%%%%%%%%%%%%%%%%%%%%%%%%
\subsubsection{Approximately Euclidean regime} \label{Euclreg}
%%%%%%%%%%%%%%%%%%%%%%%%%%%%%%%%%%%%%%%%%%%%%
We begin by solving the EOM \eqref{FriedmannEQ}-\eqref{scalarfieldEQ} subject to NB ``initial'' conditions in a regime around $z = 0$, for arbitrary $\phi_0 \in \C$. We generalize the notation of \S\ref{dSMSP} to $$H(\phi) \equiv \sqrt{V(\phi)/3} \,,$$ and define a subscript `0' on a variable to mean that it is evaluated at $\phi_0$. Expanding $a(z)$ and $\phi(z)$ in powers of $z$ and solving the equations order by order reveals the following structure:
\begin{align}
	\pm a(z;\phi_0) H_0 &= \sin \left( i H_0 z \right) - \frac{9 \varepsilon_{1,0}}{160} (i H_0 z)^5 \sum_{n=0}^\infty c_n (i H_0 z)^{2n} + \sum_{k=1}^\infty \varepsilon_{1,0} (i H_0 z)^{2k+5} \sum_{n=0}^\infty \mathcal{O}(\varepsilon^k) (i H_0 z)^{2n} \,, \label{aexpansionbig} \\
	\phi(z;\phi_0) &= \phi_0 + \frac{3 \sqrt{\varepsilon_{1,0}}}{4 \sqrt{2}} (i H_0 z)^2 \sum_{n=0}^\infty d_n (i H_0 z)^{2n} + \sum_{k=1}^\infty \sqrt{\varepsilon_{1,0}} (i H_0 z)^{2k+2} \sum_{n=0}^\infty \mathcal{O}'(\varepsilon^k) (i H_0 z)^{2n} \,. \label{phiexpansionbig}
\end{align}
In these equations $( c_n )$ and $( d_n )$ are sequences of rational numbers of decreasing magnitude, with $c_0 = d_0 = 1$. We will discuss these sequences in more detail below. The notation $\mathcal{O},\mathcal{O}'(\varepsilon^k)$ signifies a homogeneous polynomial of order $k$ in specific powers of the (``potential'' \cite{Liddle:1994dx}) slow-roll parameters, which we define for $n \geq 1$ as
\begin{equation} \label{epsilonn}
	\varepsilon_n \propto \left( \frac{V^{(n)}}{V} \right)^{2/n} \,,
\end{equation}
with $\varepsilon_1 \equiv (V'/V)^2/2$ in particular. The building blocks for a homogeneous polynomial $\mathcal{O}(\varepsilon^k)$ are the $\varepsilon_n^{n/2}$ with $1 \leq n \leq k+1$, and all possible combinations can appear except those with no factors of $\sqrt{\varepsilon_1}$ and factors other than $\varepsilon_2$. For example, $\mathcal{O}(\varepsilon^2)$ can contain terms $\propto \varepsilon_1^2, \varepsilon_1 \varepsilon_2, \varepsilon_2^2$ and $\sqrt{\varepsilon_1} \, \varepsilon_3^{3/2}$ and $\mathcal{O}(\varepsilon^3)$ can contain terms $\propto \varepsilon_1^3, \varepsilon_1^2 \varepsilon_2, \varepsilon_1 \varepsilon_2^2, \varepsilon_2^3, \varepsilon_1^{3/2} \varepsilon_3^{3/2}, \sqrt{\varepsilon_1} \varepsilon_2 \varepsilon_3^{3/2}$ and $\varepsilon_1 \varepsilon_4^2$. All of these factors are to be evaluated at $\phi_0$ -- we suppressed an additional `0' index for simplicity of notation.

As discussed in \S\ref{QCsec}, we may pick a sign for $a$ in \eqref{aexpansionbig} without loss of generality since the other solutions are related by symmetry. We will choose the $-$ sign, and additionally search for solutions in the quadrant $\Re(z_1), \Im(z_1) > 0$.

Regarding the $c_n$ and $d_n$, the first few are given by
\begin{align}
	( c_n ) &= \left( 1, -\frac{1}{42}, \frac{8}{1701}, \frac{1207}{2993760}, \frac{49099}{934053120}, \frac{1257661}{196151155200}, \cdots \right) \,, \notag \\
	( d_n ) &= \left( 1, \frac{1}{12}, \frac{7}{720}, \frac{47}{40320}, \frac{251}{1814400}, \frac{553}{34214400}, \cdots \right) \,. \notag
\end{align}
Further elements of both sequences rapidly decrease in magnitude. Numerically we found the following estimates to be accurate:
\begin{align}
	c_{n \geq 3} &\approx \exp \left( -1.37 - 2.12 n \right) \,, \notag \\
	d_{n \geq 1} &\approx \exp \left( -.293 - 2.16 n \right) \equiv A \, e^{-Bn} \,. \label{dnseries}
\end{align}
Using this last approximation we have
\begin{equation} \label{seriesphiapprox}
	\sum_{n=0}^\infty d_n (i H_0 z)^{2n} \approx 1 + A \, e^{-B} \frac{(i H_0 z)^2}{1 - e^{-B} (i H_0 z)^2} \,.
\end{equation}
According to \eqref{dnseries} the radius of convergence of this series is about $|H_0 z| \lesssim e^{B/2} \approx 2.94$, and \eqref{seriesphiapprox} is accurate well-within this disk. There is an analogous approximation for $\sum_{n=0}^\infty c_n (i H_0 z)^{2n}$ and this series has a similar radius of convergence. These precise approximations are not central to our argument though: the key point is that the radii of convergence of the two series lie close to 3 in $H_0 z$. We will return to this point later (see \cite{radius}).

To proceed we will make some assumptions on $\phi_0$, which we will justify in \S\ref{Lorreg}. These assumptions will limit the $(b,\chi)$ that we can reach with the instantons we will have constructed as we alluded to at the beginning of this section. The first two assumptions are that the first two slow-roll parameters are small at $\phioRe$:
\begin{equation} \label{SRregime}
	\varepsilon_1(\phioRe), |\varepsilon_2(\phioRe)| \ll 1 \,.
\end{equation}
The other conditions are\footnote{We will assume $\varepsilon_1(\phioRe) \neq 0$.}
\begin{align}
\left| \sqrt{\varepsilon_1(\phioRe)} \, \phi_0^\Im \right| &\ll 1 \,, \label{assumpz1} \\
\left| \varepsilon_n(\phioRe)^{n/2} (\phi_0^\Im)^{n-1} \right| &\ll \left| \sqrt{\varepsilon_1(\phioRe)} \right| \,, ~~ \forall n \geq 2 \,.\footnotemark \label{assumpz2}
\end{align}
\footnotetext{In Eq. \eqref{epsilonn} we defined $\varepsilon_n$ up to an $n$-dependent factor. These factors, together with factors that appear in the expansion of $\varepsilon_1(\phi_0)$ around $\phi_0^\Im = 0$ (see below), can become substantial at large $n$ and should ideally appear in all the inequalities we write involving $\varepsilon_n$. Determining these goes beyond the scope of this paper. However, if we define the $\varepsilon_n$ via \eqref{epsilonn} with proportionality constants equal to one, we found the inequalities we have written are sufficient.}
These last conditions ensure that
\begin{align}
	H_0 &= H(\phioRe) \left[ 1 + \mathcal{O}\left( \sqrt{\varepsilon_1(\phioRe)} \, \phi_0^\Im \right) \right] \,, \label{H0Re} \\
	\varepsilon_{1,0} &= \varepsilon_1(\phi_0) = \varepsilon_1(\phioRe) \left[ 1 + o(1) \right] \,, \label{eps0}
\end{align}
as can be seen from an expansion of these quantities around $\phi_0^\Im = 0$. In an abuse of notation we will abbreviate $H(\phioRe) \equiv \HoRe, \varepsilon_n(\phioRe) \equiv \varepsilon_{n,0}^\Re$ in the following.

To proceed further, we would like to neglect the double-series terms in Eqns. \eqref{aexpansionbig}-\eqref{phiexpansionbig} in the regime $|H_0 z| \lesssim \mathcal{O}(2)$. For this to be justified, due to the structure of these terms we described above, it is sufficient that
\begin{align}
	\left| \varepsilon_{1,0} \right| &\ll 1 \,, \label{abseps1small} \\
	\left| (\varepsilon_{1,0})^{n-2} (\varepsilon_{n,0})^n \right| &\ll 1 \,, ~~ \forall n \geq 2 \,. \label{absepsnsmall}
\end{align}
Condition \eqref{abseps1small} is satisfied due to \eqref{SRregime}-\eqref{eps0}. We can examine the conditions \eqref{absepsnsmall} by expanding the left-hand sides around $\phi_0^\Im = 0$ and using the assumptions \eqref{SRregime}-\eqref{assumpz1}-\eqref{assumpz2} we have already made. We conclude that \eqref{absepsnsmall} would be satisfied if additionally
\begin{equation} \label{ngeq3}
	\left| (\epsoRe)^{n-2} (\varepsilon_{n,0}^\Re)^n \right| \ll 1 \,, ~~ \forall n \geq 3
\end{equation}
and if
\begin{equation} \label{phi0Imsmall}
	|\phioIm| \lesssim |\sqrt{\epsoRe}| \,.
\end{equation}
If \eqref{SRregime}-\eqref{ngeq3}-\eqref{phi0Imsmall} are satisfied, so are \eqref{assumpz1}-\eqref{assumpz2}. In summary, if
\begin{align}
|\phioIm| \lesssim |\sqrt{\epsoRe}| \,, \label{assumpt1} \\
	\epsoRe, |\varepsilon_{2,0}^\Re| &\ll 1 \,, \label{assumpt2} \\
	\left| (\epsoRe)^{n-2} (\varepsilon_{n,0}^\Re)^n \right| &\ll 1 \,, ~~ \forall n \geq 3 \,, \label{assumpt3}
\end{align}
then the NB instanton we are considering is approximately given by
\begin{align}
	a(z;\phi_0) H_0 &\approx -\sin \left( i H_0 z \right) + \frac{9 \varepsilon_{1,0}}{160} (i H_0 z)^5 \sum_{n=0}^\infty c_n (i H_0 z)^{2n} \,, \label{aexpansionapprox} \\
	\phi(z;\phi_0) &\approx \phi_0 + \frac{3 \sqrt{\varepsilon_{1,0}}}{4 \sqrt{2}} (i H_0 z)^2 \sum_{n=0}^\infty d_n (i H_0 z)^{2n} \,, \label{phiexpansionapprox}
\end{align}
in the regime $|H_0 z| \lesssim \mathcal{O}(2)$. In particular, as in \S\ref{dSMSP}, we can track the solution along the Euclidean (i.e. imaginary) axis starting from the origin, and then at a turning point $\zturn$ make a 90$^\circ$ turn onto a Lorentzian segment (i.e. parallel to the real axis). From the constant potential calculation of \S\ref{dSMSP}, and from our approximation \eqref{H0Re}, we expect $\zturn \approx i \pi/(2 \HoRe)$. We have $|H_0 \zturn| \lesssim 2$ for this guess so we can certainly trust our approximation \eqref{aexpansionapprox}-\eqref{phiexpansionapprox} until we reach this point. More precisely, we will write
\begin{equation} \label{zturn}
	\zturn \equiv \frac{i \pi}{2 \HoRe} (1+\alpha) \,,
\end{equation}
with $|\alpha(b,\chi)| \ll 1$ a function that we will approximate later. Notice that neither the metric nor the scalar field are real in the Euclidean or Lorentzian directions because $\phi_0 \notin \R$.

The turning point in \eqref{zturn} is a generalization of the turning point for the $n = 0$ saddle of \S\ref{dSMSP} to slow-roll models. One may wonder what has happened to the other $n \geq 1$ saddles displayed in Fig. \ref{fig:M1}: do these have a generalization to slow-roll models as well? Specifically, might there be other solutions at $\zturn \approx (2n+1) \times i \pi/(2 \HoRe)$? We cannot answer this question with our method i.e. the approximations \eqref{aexpansionapprox}-\eqref{phiexpansionapprox}, since the radii of convergence of the series are too small to incorporate those $\zturn$ \bibnote[radius]{\`A propos, we can argue why the radii of convergence of the series in \eqref{aexpansionapprox}-\eqref{phiexpansionapprox} should be smaller than about $\pi$ in $H_0 z$: at $H_0 z \approx i \pi$, the leading term in \eqref{aexpansionapprox} vanishes and the scale factor becomes small. At this point the scalar generically blows up according to the EOM \eqref{scalarfieldEQ} in the manner described by Hawking and Turok \cite{Hawking:1998bn}, so that corrections to the leading term in \eqref{phiexpansionapprox} must become important, which in turn influence the corrections in \eqref{aexpansionapprox}. An exception to this reasoning is the CDL instanton, where all the correction series in \eqref{aexpansionbig}-\eqref{phiexpansionbig} conspire to give a smooth and finite result. As explained in \S\ref{NBsec} we are not in this scenario, however.} and we were not able to resum them either. To address the question we sought for these other solutions numerically in a handful of specific slow-roll models, but we did not find them. We cannot rule out the possibility of other saddles with $\zturn \not\approx (2n+1) \times i \pi/(2 \HoRe)$ for $n \notin \{ -1,0 \}$, however. In any case in the following we will focus on the generalized $n = 0$ saddle (which appears in four copies as we argued in \S\ref{QCsec}), for which we found both analytic and numerical evidence.

%%%%%%%%%%%%%%%%%%%%%%%%%%%%%%%%%%%%%%%%%%%%%
\subsubsection{Approximately Lorentzian regime} \label{Lorreg}
%%%%%%%%%%%%%%%%%%%%%%%%%%%%%%%%%%%%%%%%%%%%%
We now follow the solution along the line $z = \zturn + t, t > 0$ (``Lorentzian'' direction). For $\HoRe t \lesssim 1$, \eqref{aexpansionapprox}-\eqref{phiexpansionapprox} are good approximations. Beyond this regime those approximations fail. Instead, in the regime $\HoRe t \gg 1$, we return to the two complex EOM \eqref{FriedmannEQ}-\eqref{scalarfieldEQ} for the four real functions $\phi^\Re(\zturn + t), \phi^\Im(\zturn + t), a^\Re(\zturn + t), a^\Im(\zturn + t)$ and make new assumptions, namely
\begin{align}
	(\dot{\phi}^\Re)^2 &\ll V(\phi^\Re) \,, \label{assump1} \\
	(a^\Re)^2 \, V(\phi^\Re) &\gg 1 \,, \label{assump2} \\
	|\ddot{\phi}^\Re| &\ll |V'(\phi^\Re)| \,, \label{assump3} \\
	|a^\Im| &\ll |a^\Re| \,, \label{assump4} \\
	|\dot{a}^\Im| &\ll |\dot{a}^\Re| \,, \label{assump5} \\
	|\dot{\phi}^\Im| &\ll |\dot{\phi}^\Re| \,. \label{assump6}
\end{align}
The first three are the usual slow-roll assumptions from classical cosmology (for a discussion of slow-roll in the context of classical cosmology, see \cite{PhysRevD.29.2162,PhysRevD.42.3936,Liddle:1992wi,Liddle:1994dx}) for the real parts of the scale factor and scalar field -- negligible kinetic energy, and curvature, compared to potential energy, and ``slow-roll''. The last three are new assumptions for quantum cosmology. Together they imply the following approximate equations which could be called the ``slow-roll approximation in quantum cosmology'':
\begin{align}
	\frac{\dot{a}^\Re}{a^\Re} &\approx \sqrt{\frac{V(\phi^\Re)}{3}} \equiv H^\Re \,, \label{SR1} \\
	3 H^\Re \dot{\phi}^\Re &\approx - V'(\phi^\Re) \,, \label{SR2} \\
	H^\Re \dot{a}^\Im &\approx \frac{a^\Re}{6} \left( \dot{\phi}^\Re \dot{\phi}^\Im + V'(\phi^\Re) \phi^\Im \right) + (H^\Re)^2 a^\Im \,, \label{aImeq} \\
	\ddot{\phi}^\Im + 3 H^\Re \dot{\phi}^\Im &\approx \frac{3}{a^\Re} \left( H^\Re a^\Im - \dot{a}^\Im \right) \dot{\phi}^\Re - V''(\phi^\Re) \phi^\Im \approx 0 \,. \label{phiimeq}
\end{align}
Observe that the usual slow-roll equations \eqref{SR1}-\eqref{SR2} for $(a^\Re,\phi^\Re)$ can be solved independently from Eqns. \eqref{aImeq}-\eqref{phiimeq}, so that the latter can be viewed as equations for only $(a^\Im,\phi^\Im)$ after having solved \eqref{SR1}-\eqref{SR2}. The approximate equations \eqref{SR1}-\eqref{SR2} are consistent for $t > t_*$ with the assumptions \eqref{assump1}-\eqref{assump3} and thus with the exact EOM as long as
\begin{align}
	\varepsilon_1(\phi^\Re), |\varepsilon_2(\phi^\Re)| \ll 1 \,, \label{SRconsist1} \\
	a^\Re(\zturn + t_*)^2 \, V(\phi^\Re) (\zturn + t_*) \gg 1 \,. \label{SRconsist2}
\end{align}
Then, in Eq. \eqref{phiimeq}, we have neglected the terms on the right-hand side compared to those on the left-hand side. We show this is consistent below.

To summarize, the consistency conditions we must check are Eqns. \eqref{assumpt1}-\eqref{assumpt2}-\eqref{assumpt3}-\eqref{SRconsist1}-\eqref{SRconsist2}-\eqref{assump4}-\eqref{assump5}-\eqref{assump6} and the additional approximation in \eqref{phiimeq}.

We start by solving \eqref{phiimeq} under the assumption that the right-hand side can be neglected:
\begin{equation} \label{largetsolphi}
	\phi^\Im(\zturn + t) = c_1 + c_2 \int_0^t \frac{\di t'}{a^\Re(\zturn + t')^3} \,.
\end{equation}
Recall this formula is supposedly only valid for $\HoRe t \gg 1$, where it tells us that $|\phi^\Im|$ is monotonically decreasing. This is consistent with the late-time boundary condition that $\phi^\Im$ should vanish. Now, \textit{we will assume} our approximate solution \eqref{phiexpansionapprox}, valid for $\HoRe t \lesssim 1$, smoothly connects onto the solution \eqref{largetsolphi}, supposedly valid for $\HoRe t \gg 1$ (and we will assume the analogous connection for $a$). We will not be able to describe the solution in the transitional regime (in particular, unfortunately, we will not be able to prove that our ``solution'' always exists), but we will not need this detailed information for our main physical purpose namely the approximation of the action of the solution -- assuming it exists -- which determines the semiclassical wave function (see \S\ref{actionsec}). Before continuing we stress that while \eqref{aexpansionapprox}-\eqref{phiexpansionapprox} are always valid (given our assumptions \eqref{assumpt1}-\eqref{assumpt3} on $\phi_0$), the smooth connection onto a solution to the complexified slow-roll equations \eqref{SR1}-\eqref{phiimeq} is \textit{not} always valid. The connection will only be realized for a specific choice of $\phi_0$ and $\alpha$ (recall Eq. \eqref{zturn}), to which we return below. In particular the NB instanton, like the CDL instanton but unlike the classical single field slow-roll solution, is by no means an attractor. On the contrary, a general small perturbation in the parameters $\phi_0 \in \C$ or $\alpha \in \R$ destroys the solution.

Assuming a smooth connection between \eqref{phiexpansionapprox} and \eqref{largetsolphi}, and assuming the boundary condition $\phi^\Im(\zturn + t_1) = 0$ for some $t_1 \gg (\HoRe)^{-1}$, we have
\begin{align}
	c_1 &= \mathcal{O}(1) \times \sqrt{\epsoRe} \,, \notag \\
	c_2 &= - \mathcal{O}'(1) \times \left( \frac{\sqrt{\epsoRe}}{(\HoRe)^2} \right) \,, \notag
\end{align}
where $\mathcal{O}(1),\mathcal{O}'(1)$ are two positive order one numbers that depend on the details of the transitional regime $\HoRe t = \mathcal{O}(1)$. It follows that $\phi_0^\Im = \mathcal{O}(\sqrt{\epsoRe})$ for this kind of solution. We could estimate the values of the order one numbers more precisely by matching \eqref{largetsolphi} to \eqref{phiexpansionapprox} at $t = 0$ (which is invalid strictly speaking\footnote{We could also have matched at $\HoRe t = \mathcal{O}(1)$, which would appear more consistent, but this gives a very similar result to \eqref{phi0Imapprox}.}), i.e. at the turning point, to obtain
\begin{equation}
	\dot{\phi}^\Im(\zturn) \approx -2.82 \, \HoRe \sqrt{\epsoRe} = \frac{c_2}{a^\Re(\zturn)^3} \approx (\HoRe)^3 c_2 \,, \notag
\end{equation}
so that $c_2 \approx -2.82 \, \sqrt{\epsoRe}/(\HoRe)^2$. For $c_1$ we use only the zeroth order term in the slow-roll parameters in Eq. \eqref{aexpansionapprox} (this is OK because the dominant contribution to the integral in Eq. \eqref{largetsolphi} comes from early times), and equate
\begin{equation} \notag
	0 = \phi^\Im(\zturn + T) \approx \phi^\Im(\zturn + \infty) = c_1 + c_2 (\HoRe)^2 \frac{\pi}{4}
\end{equation}
so that $c_1 \approx 2.21 \sqrt{\epsoRe} \approx \phi^\Im(\zturn)$. From \eqref{phiexpansionapprox}, it follows that
\begin{equation} \label{phi0Imapprox}
	\phi_0^\Im \approx \phi^\Im(\zturn) \approx 2.21 \sqrt{\epsoRe} \,.
\end{equation}
This value for $\phi_0^\Im$ is consistent with our assumption \eqref{assumpt1}. Further, \\ $|\dot{\phi}^\Im| = \mathcal{O} \left[ \sqrt{\epsoRe} (\HoRe)^{-2} (a^\Re)^{-3} \right]$ while $|\dot{\phi}^\Re| = \mathcal{O} \left( \sqrt{\varepsilon_1^\Re} \, H^\Re \right)$ so that \eqref{assump6} is satisfied.

The solution to \eqref{aImeq} is approximately given by
\begin{equation} \label{aImlate}
	a^\Im \approx - \frac{\sqrt{2}}{3} \sqrt{\varepsilon_1^\Re} 
	\, a^\Re \phi^\Im
\end{equation}
at late times $\HoRe t \gg 1$. To see this note that
\begin{equation} \label{phiImlate}
	\phi^\Im \approx - \frac{c_2}{3 H^\Re (a^\Re)^3}
\end{equation}
in the late-time regime, which can be seen from \eqref{largetsolphi} by changing variables $t' \rightarrow a^\Re$, using \eqref{SR1} and that the fractional change $| \di \log (1/H^\Re)/\di \log (a^\Re)^{-3} | \propto \varepsilon_1^\Re$ (which follows from \eqref{SR1}-\eqref{SR2}), which we assume is much smaller than unity. Alternatively \eqref{phiImlate} solves \eqref{phiimeq} up to corrections involving the slow-roll parameters. With this solution the reader may verify the consistency conditions \eqref{assump4}-\eqref{assump5}, and justify our neglect of the right-hand side in Eq. \eqref{phiimeq}.

With the knowledge of $a^\Im$ we can return to the approximation of $\alpha$ in Eq. \eqref{zturn}. Assuming a smooth connection between \eqref{aImlate} and \eqref{aexpansionapprox}, we have $|\dot{a}^\Im(\zturn)| = \mathcal{O}\left( \epsoRe \right)$, from which it can be seen that $|\alpha| =\mathcal{O}(1) \times \epsoRe$. To estimate the order one constant more precisely, as we did in \eqref{phi0Imapprox}, we can try to match \eqref{aImlate} and \eqref{aexpansionapprox} at some intermediate values $\HoRe t = \mathcal{O}(1)$. This gives
\begin{equation} \label{alpha0eq}
	\alpha \approx -0.05 \times \epsoRe \,.
\end{equation}

Finally we turn to $a^\Re$ and $\phi^\Re$. At late times $\HoRe t \gg 1$ these follow the classical slow-roll evolution i.e. Eqns. \eqref{SR1}-\eqref{SR2}, as long as $\varepsilon_1^\Re, |\varepsilon_2^\Re| \ll 1$. The remaining question is what the initial conditions are, say at some intermediate $\HoRe t_* = \mathcal{O}(1)$ when the slow-roll approximations \eqref{SR1}-\eqref{phiimeq} first become accurate. We will write $a^\Re_* \equiv a^\Re(t_*)$, $\phi^\Re_* \equiv \phi^\Re(t_*)$ to indicate the values of the scale factor and scalar field at the onset of the inflationary period. From \eqref{phiexpansionapprox} we infer 
\begin{equation} \notag
	\phi^\Re_* = \phioRe \pm \mathcal{O}(\sqrt{\epsoRe})\approx \phioRe \,,
\end{equation}
since $\phi^\Re$ goes from $\phioRe$ to $\phioRe + \mathcal{O}(1) \times \sqrt{\epsoRe}$ during the approximately Euclidean phase, and from $\phioRe + \mathcal{O}(1) \times \sqrt{\epsoRe}$ to $\phioRe + \left[ \mathcal{O}(1) - \mathcal{O}'(1) \right] \times \sqrt{\epsoRe}$ during the initial approximately Lorentzian phase, where $\mathcal{O}(1),\mathcal{O}'(1)$ are two positive order one numbers. For $a^\Re$ we have
\begin{equation} \notag
	a^\Re_* = \frac{\mathcal{O}(1)}{\HoRe}
\end{equation}
where $\mathcal{O}(1) \gtrsim 1$. This last result validates our assumption \eqref{SRconsist2}. There remain the conditions \eqref{SRconsist1} on the real part of the scalar trajectory (namely, that it remains in a slow-roll patch) and the conditions \eqref{assumpt2}-\eqref{assumpt3} on $\phioRe$ (essentially, that the trajectory starts out in a slow-roll patch).

From the slow-roll equations \eqref{SR1}-\eqref{SR2}, and from the late-time boundary condition $a(\zturn + t_1) = b, \phi(\zturn + t_1) = \chi$, we deduce the following relationship between $\phioRe, b$ and $\chi$ provided $b H(\chi) \gg 1$ and $\varepsilon_1(\chi), |\varepsilon_2(\chi)| \ll 1$:
\begin{equation} \label{phi0bchirelation}
	b \approx a_* \, \exp \left( \int_{\chi}^{\phi_*} \di \phi \, \frac{V(\phi)}{V'(\phi)} \right) \,.
\end{equation}
This relation makes more precise the comment we made at the beginning of \S\ref{SRsec}, namely that our method only describes a NB instanton which reaches particular values of $(b,\chi) \in \R^+ \times \R$. The values $(b,\chi)$ are reached by an instanton we have described if they lie on a slow-roll trajectory $(a^\Re,\phi^\Re)$ which has passed through $(a^\Re_*,\phi_*^\Re) = \left( \mathcal{O}(1)/\HoRe, \phioRe \right)$ for some $\phioRe$ where the slow-roll parameters are small (in particular, the slow-roll parameters must be small at $\chi$ as well and curvature must be negligible).

%%%%%%%%%%%%%%%%%%%%%%%%%%%%%%%%%%%%%%%%%%%%%
\subsubsection{Action} \label{actionsec}
%%%%%%%%%%%%%%%%%%%%%%%%%%%%%%%%%%%%%%%%%%%%%
To compute the action we use Eq. \eqref{Sbchicontour} and choose the contour of integration to first run from the origin along the imaginary axis until $\zturn$ (using the approximate form of the solution we found in \S\ref{Euclreg}) and then to run parallel to the real axis until the point $z_1 = \zturn+t_1$ where the arguments $(b,\chi)$ of the wave function are attained (using the approximate form of the solution we found in \S\ref{Lorreg}). We denote these two contributions to $S$ by $S_\textsf{E}$ and $S_\textsf{L}$ respectively. A short computation shows
\begin{equation} \label{SEint}
	S_\textsf{E} = \mathcal{O} \left( \frac{\epsoRe}{(\HoRe)^2} \right) + \frac{2i}{(\HoRe)^2} \left[ 1 + \mathcal{O} \left( \epsoRe \right) \right] \,.
\end{equation}
To approximate $S_\textsf{L}$ we may expand the integrand in \eqref{Sbchicontour} to first order in $a^\Im$ and $\phi^\Im$ -- that higher order terms are subdominant follows from our analysis of the solution in \S\ref{Lorreg}. We obtain
\begin{equation} \label{SLint}
	S_\textsf{L} \approx -2 \int_0^{t_1} \di t ~ a^\Re \left[ (a^\Re)^2 \, V(\phi^\Re) - 3 \right] + i \left[ 3 V(\phi^\Re) a^\Im (a^\Re)^2 + V'(\phi^\Re) \phi^\Im (a^\Re)^3 - 3 a^\Im \right] \,.
\end{equation}
To approximate the real part we use that $(a^\Re)^2 \, V(\phi^\Re) \gg 1$ during the bulk of the range of integration. This gives
\begin{equation} \notag
	S_\textsf{L}^\Re \approx -2 \int_{a_*^3}^{b^3} \di (a^\Re)^3 ~ H^\Re \,,
\end{equation}
where we also changed variables and used the slow-roll equations \eqref{SR1}-\eqref{SR2}. As we noted before, the fractional change $| \di \log H^\Re /\di \log (a^\Re)^{3} | \propto \varepsilon_1^\Re$ is small throughout the range of integration, so that the dominant contribution to the integral comes from the region around the endpoint where the scale factor is large. This gives
\begin{equation} \notag
	S^\Re_\textsf{L} \approx -2 \, b^3 H(\chi) \,.
\end{equation}
The imaginary part of the integrand in expression \eqref{SLint} is of order $\epsoRe/\HoRe$ at early times $\HoRe t \lesssim \mathcal{O}(1)$, and so the contribution to $S_\textsf{L}^\Im$ from these times is of order $\epsoRe/(\HoRe)^2$, which is subdominant compared to the imaginary part of the contribution $S_\textsf{E}$ we computed in Eq. \eqref{SEint}. At late times $\HoRe t \gg 1$ the contribution from the $-3 a^\Im$ term is negligible because this term rapidly decays. The other two terms do not separately decay, but their sum does due to the relation \eqref{aImlate}. So also this contribution to $S^\Im$ is subdominant compared to the one we have already computed. Putting everything together we conclude
\begin{equation} \label{Sbchi}
	S(b,\chi) \approx -2 \, b^3 H(\chi) + \frac{2i}{(\HoRe)^2} \,,
\end{equation}
where recall $\phioRe, b$ and $\chi$ are connected via Eq. \eqref{phi0bchirelation}. The three other NB instantons related by parity and complex conjugation operations have actions $S^*, -S$ and $-S^*$ (\S\ref{QCsec}). As a consistency check the reader may verify that the semiclassical WDW equation \eqref{HJeqaphi} is indeed satisfied by this action, to leading order in the slow-roll parameters. For this it is convenient to use Eqns. \eqref{derv1}-\eqref{derv2} below.

The saddle with action \eqref{Sbchi} (or with $-S^*$) determines the Vilenkin wave function\footnote{In \cite{Vilenkin1984} Vilenkin proposes a definition of his wave function that involves a ``Lorentzian path integral'', meaning an integral of $e^{iS/\hbar}$ over \textit{real} fields on a compact manifold. On general grounds this integral is not expected to diverge exponentially in the limit $\hbar \rightarrow 0$, which determines the sign of $S^\Im$.}, giving
\begin{equation} \label{VilenkinWF}
	\Psi_\textsf{V}(b,\chi) \approx P \, \exp\left( \pm 2i b^3 H(\chi)/\hbar \right) \exp\left( -\frac{2}{\hbar (\HoRe)^2} \right) ~~ \text{as } \hbar \rightarrow 0 \,,
\end{equation}
while the Hartle-Hawking wave function is determined by the saddles with $-S$ and $S^*$, giving
\begin{equation} \label{HHWF}
	\Psi_\textsf{HH}(b,\chi) \approx P' \, \exp\left( 2i b^3 H(\chi)/\hbar \right) \exp\left( \frac{2}{\hbar (\HoRe)^2} \right) + \text{c.c.} ~~ \text{as } \hbar \rightarrow 0 \,,
\end{equation}
where $P,P'(b,\chi)$ are prefactors that our minisuperspace analysis cannot capture.\footnote{They depend on the fluctuations of all the degrees of freedom in the full theory around the saddles -- a computation we do not attempt here (but again see \cite{Barvinsky:1990ga,Barvinsky:1992dz}, or \cite{Saad:2019lba,Maldacena:2019cbz,Godet:2020xpk} for exact results in 2D gravity models).}

%%%%%%%%%%%%%%%%%%%%%%%%%%%%%%%%%%%%%%%%%%%%%
\subsubsection{Classical histories} \label{classicalsec}
%%%%%%%%%%%%%%%%%%%%%%%%%%%%%%%%%%%%%%%%%%%%%
We now turn to the classicality condition \eqref{classicalitycondition} for semiclassical NB wave functions $\Psi_\textsf{NB} \sim P \, e^{i S/\hbar}$ as $\hbar \rightarrow 0$ with $S$ given by \eqref{Sbchi} (or $S^*, -S$ or $-S^*$, or a sum of such terms). We first use Eq. \eqref{phi0bchirelation} to derive
\begin{equation} \notag
	\partial_b \phioRe \approx \frac{\sqrt{2 \, \epsoRe}}{b} \,, ~~
	\partial_\chi \phioRe \approx \sqrt{\frac{\epsoRe}{\varepsilon_1(\chi)}} \,.
\end{equation}
It follows that
\begin{align}
	\partial_b S^\Im &\approx - \frac{4 \, \epsoRe}{(\HoRe)^2 b} \,, &
	\partial_\chi S^\Im &\approx \frac{-2 \sqrt{2} \, \epsoRe}{(\HoRe)^2 \sqrt{\varepsilon_1(\chi)}} \,, \label{derv1} \\
	\partial_b S^\Re &\approx -6 \, b^2 H(\chi) \,, &
	\partial_\chi S^\Re &\approx -\sqrt{2} \, b^3 H(\chi) \sqrt{\varepsilon_1(\chi)} \,, \label{derv2}
\end{align}
so that the classicality condition is satisfied in the region of minisuperspace where our approximation \eqref{Sbchi} holds (the additional classicality conditions of Ref. \cite{Hartle:2008ng} are also satisfied in the $(b,\chi)$-basis, but see footnote \ref{classftnote}). The classical histories \eqref{classicaluniverses} (approximately) satisfy
\begin{align}
	\dot{a}_\textsf{cl} &= a_\textsf{cl} H(\phi_\textsf{cl}) \,, \notag \\
	3 H(\phi_\textsf{cl}) \dot{\phi}_\textsf{cl} &= -V'(\phi_\textsf{cl}) \,, \notag
\end{align}
that is, the usual slow-roll equations from classical cosmology (and they pass through the point $(a_\textsf{cl},\phi_\textsf{cl}) = (b,\chi)$). Since the real part $(a^\Re,\phi^\Re)$ of the instanton (approximately) satisfies these equations at late times $1 \ll \HoRe t \lesssim \HoRe t_1$ and passes through $(b,\chi)$, we conclude that the predicted classical history and the real part of the instanton (approximately) coincide in this regime (though these two are different from each other in general). The classical histories can be labeled by the value of the scalar at the time when the universe had size $a_\textsf{cl} H_\textsf{cl} \approx \mathcal{O}(1)$, that is, by $\phioRe$. They receive a probability $\propto \exp\left[ \pm 4/ \hbar \, (\HoRe)^2 \right]$ depending on which kind of saddle dominates the wave function in the semiclassical limit.

%%%%%%%%%%%%%%%%%%%%%%%%%%%%%%%%%%%%%%%%%%%%%
\section{Discussion} \label{discussionsec}
%%%%%%%%%%%%%%%%%%%%%%%%%%%%%%%%%%%%%%%%%%%%%

%%%%%%%%%%%%%%%%%%%%%%%%%%%%%%%%%%%%%%%%%%%%%
\subsubsection*{Regime of validity of our result}
%%%%%%%%%%%%%%%%%%%%%%%%%%%%%%%%%%%%%%%%%%%%%
We repeat that our computation of the action \eqref{Sbchi}, and thus of the semiclassical approximation to a NB wave function (like the Vilenkin or Hartle-Hawking wave functions \eqref{VilenkinWF}-\eqref{HHWF}), is only valid in a limited subset of the minisuperspace, namely the regime $\{ (b,\chi) \}$ where the entire trajectory which follows the classical slow-roll evolution \eqref{SR1}-\eqref{SR2}, ending at $(a,\phi)_1 = (b,\chi)$ and starting at $(a,\phi)_0 = \left( \mathcal{O}(1)/H(\phioRe), \phioRe \right)$, is consistent (meaning the slow-roll parameters $\varepsilon_1,\varepsilon_2$ and the curvature $(a H)^{-2}$ are small along the entire trajectory).\footnote{We remind the reader that we defined $H(\phi) \equiv \sqrt{V(\phi)/3}$ in this paper.} We make no general claims about the behavior of a NB wave function outside this regime, e.g. about the behavior in the Euclidean regime $b H < 1$ (see \cite{PhysRevD.28.2960,PhysRevD.37.888}), an initial singularity or bounce in the predicted classical histories, the exit of inflation or the asymptotic future of the universe (see \cite{Hartle:2008ng}). Note that our result holds in particular for small field inflation models, which may be the only consistent effective field theories of inflation \cite{Ooguri:2006in}. Finally, we emphasize that our construction assumes two solutions (approximately Euclidean and Lorentzian respectively) may be matched onto each other in a transitional regime which lies at the border of the regimes of validity of both solutions. In all the cases we examined numerically this can be done, and while we have analytically deduced the rather precise consistency conditions \eqref{phi0Imapprox} and \eqref{alpha0eq} on such a matching, we were not able to rigorously prove that the matching can always be done.

%%%%%%%%%%%%%%%%%%%%%%%%%%%%%%%%%%%%%%%%%%%%%
\subsubsection*{Measure on the amount of inflation}
%%%%%%%%%%%%%%%%%%%%%%%%%%%%%%%%%%%%%%%%%%%%%
In \S\ref{classicalsec} we showed that in single field inflation models a NB wave function predicts a one-parameter family of classical slow-roll histories which can be labeled by the value of the scalar $\phioRe$ when the universe had size $a H \sim 1$. Without any further measure factors\footnote{\label{measureftn}Such as a volume factor that favors large universes, which may radically alter the probability \cite{Hartle:2010vi,Hartle:2013oda,Hertog:2015zwh,Hartle:2016tpo}.}, which we will not expand upon here, the probability for each history scales either as $\exp\left[ -12 / \hbar \, V(\phioRe) \right]$ (Vilenkin wave function) or $\exp\left[ 12 / \hbar \, V(\phioRe) \right]$ (Hartle-Hawking wave function). The former choice favors a high starting point and many efolds of inflation, though, how many precisely is unclear because our approximations fail at the cutoff of the theory. The latter choice favors a small amount of inflation (and is ruled out without further measure factors, but see footnote \ref{measureftn}), though, again, how small exactly is unclear because our approximations fail near the low energy threshold of inflation.\footnote{\label{HHHfnt}For example, in the $V = \Lambda + m^2 \phi^2/2$ model with $m^2 > 3 \Lambda/4$, the Hartle-Hawking state predicts that large classical universes underwent a \textit{minimum} ($N_e = \mathcal{O}(1)$, however) amount of scalar field-driven inflation in the past \cite{Hartle:2008ng}.} In any case quantum cosmology yields a very specific measure on the initial conditions for classical cosmology. This small set of preferred initial conditions typically leads to very specific predictions for cosmological observables \cite{Hertog:2013mra,Hartle:2013oda,Hertog:2015zwh} implying a high degree of falsifiability for the combination (theory of initial conditions)+(inflationary model), which is appealing. This should be contrasted with approaches where one claims to assess the predictivity of an inflationary model by drawing initial conditions from a broadly distributed (ad hoc) prior and verifying whether the resulting cosmological observables including $n_s, \alpha$ and $r$ are narrowly (``predictive'') or broadly (``not predictive'') distributed (e.g. \cite{Frazer:2013zoa,Easther:2013rva,Price:2014ufa}).

Finally we stress an important limitation of our work following from our inability to compute the wave function -- for general $V(\phi)$ -- in the entire minisuperspace (see the \textbf{Regime of validity of our result} section above). Since we were only able to compute the wave function in a (general) slow-roll regime, we could only (generally) find the relative probability of slow-roll histories with varying amounts of inflation. We could not, for instance, find the probability to have had inflation in the past vs. to not have had inflation in the past. As we've already mentioned in footnote \ref{HHHfnt}, a study which goes beyond this restriction (albeit in a particular model) is \cite{Hartle:2008ng}: in a quadratic minimally coupled scalar field model with a cosmological constant which is small compared to the mass of the field, one concludes that all universes which behave classically at late times according to a no-boundary wave function, underwent a minimum period of inflation at early times.\footnote{The classicality conditions, which we believe merit critical examination as we mentioned in footnote \ref{classftnote}, are crucial to this argument.} It would be interesting to understand whether this argument can be generalized to other models.

%%%%%%%%%%%%%%%%%%%%%%%%%%%%%%%%%%%%%%%%%%%%%
\subsubsection*{Comparison with the literature}
%%%%%%%%%%%%%%%%%%%%%%%%%%%%%%%%%%%%%%%%%%%%%
Though we use a different method, our results agree with and generalize part of the analysis in \cite{PhysRevD.46.1546} (who studied the $V = m^2 \phi^2$ case) to arbitrary slow-roll models, and provide quantitative evidence for several claims made in \cite{Hartle:2008ng} (including the anticipation of the result \eqref{HHWF}). We offer three critical thoughts on statements made in \cite{PhysRevD.46.1546}, however.

The first regards the discussion about the late-time future of the classical histories in the $m^2 \phi^2$ model (we mentioned above in \textbf{Regime of validity of our result} that this part of the minisuperspace lies beyond the scope of our calculation). \cite{PhysRevD.46.1546} argues that because $a^\Im$ and $\phi^\Im$ are ``approximately zero'' after some period of inflation (indeed we showed in \S\ref{Lorreg} that these functions decay as $H(\phi^\Re)^{-1} H(\phioRe)^{-2} (a^\Re)^{-2}$ and $H(\phi^\Re)^{-1} H(\phioRe)^{-2} (a^\Re)^{-3}$ during inflation respectively), they will remain this way after inflation has ended. This is unclear. For example, if we consider the evolution of a NB instanton along the Lorentzian line $\zturn + t$, for which $a(\zturn + t_1) = b, \phi(\zturn+t_1) = \chi$ for some $H(\phioRe) t_1 \gg 1$, we found that the solution does \textit{not} remain approximately real for times $t \gg t_1$ (this is a numerical result that we found to be true in each case we examined, including in the $V = m^2 \phi^2$ model). Instead, e.g., $a^\Im$ diverges away from its decaying path and follows $a^\Im \sim \pm a^\Re$. This result does not invalidate the claim that there is a (na\"ive\footnote{Large quantum fluctuations in an eternally expanding universe prohibit it from being eternally classical.}) classical regime in the infinite volume limit -- we are merely pointing out that neither our current analysis nor the one in \cite{PhysRevD.46.1546} provides evidence for it.

The second comment regards the claimed existence of specific other $O(4)$-symmetric saddles on $B^4$, namely generalizations of the $n \notin \{ 0,1 \}$ saddles of Fig. \ref{fig:M2} (constant potential) to slow-roll models (this point was mentioned in \S\ref{Euclreg}). As we mentioned, we found neither numerical nor analytic evidence for such saddles.

The final comment regards the calculation of $S^\Im$ (our Eq. \eqref{Sbchi}). Ref. \cite{PhysRevD.46.1546} finds $S^\Im \propto (\HoRe)^{-2}$ (applied to the quadratic model), but was not able to find the proportionality constant. Our detailed calculation determines the constant, which indeed coincides with the known constant in the case of a constant potential. \\

\noindent Then we point out a potentially imprecise equation that one can find in the literature, e.g. in \cite{PhysRevD.37.888,Kiefer2012}, namely
\begin{align} \label{incorrecteq1}
	S(b,\chi) \approx -2 \frac{\left[ (b H(\chi))^2 - 1 \right]^{3/2}}{H(\chi)^2}& + \frac{2i}{H(\chi)^2} \,, ~~~ \text{when } b H(\chi) > 1 \text{ and } \varepsilon_1(\chi) \ll 1 \,. \notag \\
	&\text{(an equation inferred from \cite{PhysRevD.37.888})}
\end{align}
If we assume $b H(\chi) \gg 1$ and $\varepsilon_1(\chi) \ll 1$, which are less restrictive than the conditions for the validity of our result in Eq. \eqref{Sbchi} (see the paragraph on \textbf{Regime of validity of our result} above), this becomes
\begin{equation} \label{incorrecteq}
	S(b,\chi) \approx -2 \, b^3 H(\chi) + \frac{2i}{H(\chi)^2} \,. \hspace{2cm} \text{(potentially imprecise)}
\end{equation}
This approximation differs from our result \eqref{Sbchi} when $H(\chi) \not\approx \HoRe \equiv H(\chi_0)$, where $\chi_0$ is the value of the scalar at the start of inflation. If inflation lasts long enough, and if the potential is suitable, $H(\chi)$ may become arbitrarily different from $H(\chi_0)$. In this regime one can verify that \eqref{incorrecteq} does not approximately satisfy the WDW equation while the expression \eqref{Sbchi} does. One way to see this is from the imaginary part of Eq. \eqref{HJeq}, $\partial S^\Re \cdot \partial S^\Im = 0$. In the proposal \eqref{incorrecteq}, $S^\Im$ does not depend on $b$ so the equation becomes $\partial_\chi S^\Re \,  \partial_\chi S^\Im = 0$, which is not (approximately) satisfied. Instead in \eqref{Sbchi}, $S^\Im$ depends on $b$ (and $\chi$) in a particular way. $\partial S^\Re \cdot \partial S^\Im$ now becomes a sum of two terms, which, one can verify, is subleading in slow-roll compared to the two terms separately. The equation has been approximately solved. Another way to see that \eqref{incorrecteq} is imprecise in this regime is by noting that $S^\Im$ must be (exactly) constant along the integral curves of $S^\Re$ (as we reviewed in \S\ref{QCsec}). As we discussed in \S\ref{classicalsec}, these integral curves are the slow-roll solutions. Though $H(\chi)$ is slowly varying during slow-roll, it is not exactly constant, and indeed, as we mentioned, it may change by an arbitrary amount in general. By contrast $H(\chi_0)$ is, of course, constant. The expression \eqref{Sbchi} generalizes the result \eqref{iSdSMSP} in the case of a constant potential (again, in the appropriate classical regime, see \textbf{Regime of validity of our result} above) to slow-roll models, in the sense that the real part of the action is still predominantly determined by the (approximately) Lorentzian, classical slow-roll evolution, while the imaginary part, which determines the probability of the trajectory, is determined for the most part by the (approximately) Euclidean piece.

We stress that our remark about the potential impreciseness of \eqref{incorrecteq1} pertains only to the classical regime $b H(\chi) \gg 1$. When $b H(\chi) \lesssim 1$ -- a regime we did not study in this work and where our Eq. \eqref{Sbchi} does not apply -- Eq. \eqref{incorrecteq1} is a good approximation (since the scalar has not moved much we can simply copy the constant potential result in this regime with the replacement $H \rightarrow H(\chi)$). \\

\noindent Finally we mention the recent work \cite{Matsui:2020tyd} which in the Appendix also considers NB instantons in slow-roll models. One assumes a quadratic approximation to the potential which limits the scope of the calculation (and only slightly generalizes \cite{PhysRevD.46.1546}), but, within this scope and given a modification (to be implemented in a published version \cite{corresp}), the result for $S$ agrees with our \eqref{Sbchi} to leading order in the slow-roll parameters.

%%%%%%%%%%%%%%%%%%%%%%%%%%%%%%%%%%%%%%%%%%%%%
\subsubsection*{No overshoot problem}
%%%%%%%%%%%%%%%%%%%%%%%%%%%%%%%%%%%%%%%%%%%%%
The overshoot problem is the apparent tension between two features of inflationary models that descend from string theory \cite{Brustein:1992nk}. On the one hand, for consistency reasons, the inflationary plateau must lie well-below the Planck scale. Assuming the field descends from a region of string scale energy density, it will have a large kinetic energy when it arrives at the inflationary plateau. On the other hand, also for consistency reasons, the extent of the inflationary regime is expected to be small in Planck units \cite{Ooguri:2006in}. This leads to the expectation that the inflaton will generically ``overshoot'' the inflationary plateau, and so inflation will not occur.

In \cite{Freivogel:2005vv} one explains how this problem is resolved in bubble universes that arise from a CDL tunneling event: the negative curvature inside such bubbles provides a friction term that brings the inflaton to a halt on the inflationary plateau, independently of the magnitude of the initial kinetic energy. In the NB proposal the universe is positively curved, so this mechanism is not available. Instead, it is the imaginary time evolution that provides an effective friction term.\footnote{We should be quick to note that this assumes the instanton we have identified in this paper provides the dominant contribution to the semiclassical wave function. As we mentioned in \S\ref{Euclreg} we found no evidence for other $O(4)$-symmetric solutions on $B^4$, but we are not able to exclude them altogether, let alone possibly dominant solutions on other manifolds, which could reintroduce the overshoot problem.} Because the solution is smooth, $\dot{\phi} = 0$ at the south pole of the $S^4$. At the equator, the field only has kinetic energy $\propto \varepsilon_1 H^2$ which is small compared to the potential energy $\propto H^2$. These are two mechanisms which solve the overshoot problem by making $|\dot{\phi}| \ll H$ at the onset of inflation (in e.g. \cite{Easther:2013bga,Easther:2013rva} it is claimed such a mechanism does not exist).

%%%%%%%%%%%%%%%%%%%%%%%%%%%%%%%%%%%%%%%%%%%%%
\subsubsection*{Extension to multifield models}
%%%%%%%%%%%%%%%%%%%%%%%%%%%%%%%%%%%%%%%%%%%%%
We deem it likely that our results can be generalized to the case of multiple homogeneous scalar fields that live on a curved manifold, i.e. to a non-linear sigma model, or even to more general ``$P(X)$''-theories \cite{Langlois:2008mn}. We leave an investigation of this highly relevant generalization (since many fields typically appear in string cosmology models, e.g. an $\mathcal{O}(100)$ in axion theories \cite{Arvanitaki:2009fg,Bachlechner:2019vcb}) to future work.

%%%%%%%%%%%%%%%%%%%%%%%%%%%%%%%%%%%%%%%%%%%%%
\section*{Acknowledgments}
%%%%%%%%%%%%%%%%%%%%%%%%%%%%%%%%%%%%%%%%%%%%%
Thomas Hertog, Matt Kleban and Alex Vilenkin are thanked for many discussions. This work was supported by the Carl P. Feinberg Graduate Fellowship at NYU. We thank Hiroki Matsui and Takahiro Terada for correspondence on a draft of this paper. The author expresses gratitude towards the Wynendaele family of Haasrode, Belgium, for their hospitality while this work was initiated and towards the ITF at the KU Leuven for their hospitality while this work was completed.

%%%%%%%%%%%%%%%%%%%%%%%%%%%%%%%%%%%%%%%%%%%%%
\nocite{Hawking:1998bn}
\bibliographystyle{klebphys2}
\bibliography{refs}
%%%%%%%%%%%%%%%%%%%%%%%%%%%%%%%%%%%%%%%%%%%%%
\end{document}